\documentclass[11pt]{article}

\title{\vspace{25mm} \bf{On-shell equivalence of two formulations \\ for superstring field theory} \vspace{15mm} }
\author{\Large{Keiyu Goto\footnote{goto@hep1.c.u-tokyo.ac.jp} 
, \hspace{2mm} Hiroaki Matsunaga}\footnote{hiroaki.matsunaga@yukawa.kyoto-u.ac.jp} \vspace{5mm}}
\date{${}^{\ast }$Institute of Physics, University of Tokyo, \\ Komaba, Meguro-ku, Tokyo 153-8902, Japan \\ \vspace{3mm}
${}^{\dagger }$Yukawa Institute of Theoretical Physics, Kyoto University, \\ Kyoto 606-8502, Japan \vspace{5mm}}

\usepackage[dvips]{graphicx}
\usepackage{amsmath,amscd, amssymb}
\usepackage{cite}
\usepackage{mathrsfs} 
\usepackage{amsfonts}
\usepackage[dvipdfmx, svgnames]{xcolor}%before tikz
\usepackage{tikz}

\usepackage{upgreek}

\makeatletter
  
  \@addtoreset{equation}{section}
\makeatother

\newcommand{\ld}{ [ \hspace{-0.6mm} [ }
\newcommand{\rd}{ ] \hspace{-0.6mm} ] }
\newcommand{\Ld}{ \big[ \hspace{-1.1mm} \big[ }
\newcommand{\Rd}{ \big] \hspace{-1.1mm} \big] }

\newcommand{\no}{\nonumber\\}
\newcommand{\niu}[1]{\vspace{7pt}\noindent\underline{{\sf\hspace{4pt}#1\hspace{4pt}}}\vspace{4pt}}
\newcommand{\1}{\mbox{1}\hspace{-0.25em}\mbox{l}}

\allowdisplaybreaks[3]

\begin{document}

\maketitle 

{\vspace{-128mm}
\rightline{\tt YITP/15-38}
\rightline{\tt UT-Komaba/15-2}
\vspace{128mm}}

\begin{abstract}
%WZW-type = $L_{\infty }$-type.
In this paper we derive the condition providing the on-shell equivalence of $L_\infty$-type and WZW-like formulations for superstring field theory.
We construct the NS string products ${\bf L} = \{ L_{n} \} _{n=1}^{\infty }$ of $L_{\infty }$-type formulation and the shifted BRST operator $Q_{\mathcal{G}}$ in WZW-like formulation by the similarity transformations of the BRST operator $Q$.
Utilizing the similarity transformations, we can consider a morphism connecting the $L_\infty$-algebras on both sides.
It naturally induces the field redefinitions and guarantees the equivalence of the on-shell conditions in two formulations.
In addition, we have confirmed up to quartic order that the on-shell equivalence condition also provides the off-shell equivalence.
Then partial-gauge-fixing conditions giving $L_\infty$-relations in WZW-like formulation naturally appear.
\end{abstract}

\thispagestyle{empty}

\clearpage

\tableofcontents

\setcounter{page}{1}
%\addcontentsline{toc}{part}{Introduction}

\section{Introduction}

The complete formulation for superstring field theory has been the important problem.
However, the action in the same form as well-understood bosonic string field theory\cite{Witten:1985cc, Hata:1986, Thorn:1986qj,Kugo:1989,Sonoda:1989wa,Sen:1990ff,Schubert:1991en,Zwiebach:1992ie,Gaberdiel:1997ia,Nakatsu:2001da} has some disadvantages: divergences and broken gauge invariance\cite{Witten:1986qs, Wendt:1987zh}.
To construct a gauge invariant action consistently, various approaches have been proposed\cite{Arefeva:1989cp,Preitschopf:1989fc,Saroja:1992vw,Jurco:2013qra}. 
In this paper we focus on two formulations: Wess-Zumino-Witten-like (WZW-like) formulation and $A_\infty /L_\infty$-type formulation, and discuss their equivalence.

\vspace{1mm}

The WZW-like formulation\cite{Berkovits:1995ab,Berkovits:1999bs,Berkovits:1998bt,Berkovits:2004xh,Iimori:2013kha,Matsunaga:2013mba,Matsunaga:2014wpa} is one of the successful approaches.
The theory is formulated 
by using the string field which belongs to the state space spanned by bosonized superconformal ghosts $(\xi , \eta , \phi )$: the large Hilbert space.
The resultant action has the different form that the bosonic one and is invariant under the large gauge transformation. 
The largeness of the state space and the gauge symmetry suggests this approach would be more fundamental.
On the other hand, however, it makes the construction of a classical Batalin-Vilkovisky (BV) master action complicated,
and the quantization of WZW-like theory remains mysterious. 

\vspace{1mm}

The alternative one is the $A_\infty /L_\infty$-type formulation proposed in \cite{Erler:2013xta, Erler:2014eba}.
The action consists of an infinite set of string products satisfying $A_\infty /L_\infty$-relations and 
the string field belonging to 
the subspace of the large Hilbert space annihilated by the zero-mode of $\eta$-current: the small Hilbert space. 
The $A_\infty /L_\infty$-relations arises
the nilpotent gauge invariance,
and therefore, one can carry out the classical BV quantization of the action in a straightforward way.

\vspace{2mm}

In this paper, we discuss the relation between these two formulations for NS closed string field theory. 
First, we show that the NS string products ${\bf L} = \{ L_{n} \} _{n=1}^{\infty }$ in $L_{\infty }$-type formulation and the shifted BRST operator $Q_{\mathcal{G}}$ in WZW-like formulation both are obtained by similarity transformations of the BRST operator $Q$. 
Utilizing invertible maps ${\bf G} = \{ {\sf G}_{n} \} _{n=1}^{\infty }$ and $\mathcal{E}_{V}$, one can write ${\bf L} = {\bf G} \, {\bf Q} \, {\bf G}^{\dagger }$ and $Q_{\mathcal{G}} = \mathcal{E}_{V} \, Q \, {\mathcal{E}_{V}}^{\dagger }$, where ${\sf G}_{n}$ is a $n$-fold multilinear map. 
Then, combining these invertible maps, we construct ${\sf F} = \{ {\sf F}_{n} \} _{n=1}^{\infty } := \{ \mathcal{E}_{V} {{\sf G}^{\dagger }}_{n} \} _{n=1}^{\infty }$ and obtain the similarity transformation connecting $Q_{\mathcal{G}}$ and ${\bf L}$
\begin{align}
{\sf F} \, {\bf L} \, {\sf F}^{\dagger } = Q_{\mathcal{G}} . 
\end{align}
Second, we consider a morphism of $L_{\infty }$-algebras which is given by the similarity transformation. 
We also show that there exists a natural field redefinition induced by the $L_{\infty }$-morphism ${\sf F}$ 
\begin{align}
\Phi ^{\prime }  := \sum_{n=1}^{\infty } \frac{1}{n!} {\sf F}_{n} (\overbrace{\Phi , \dots ,\Phi }^{n} ) , 
\end{align} 
which is the key ingredient to discuss the on-shell equivalence. 
Since the $L_{\infty }$-morphism ${\sf F}$ maps the state ${\bf L} (e^{\wedge \Phi })$ to the state $Q_{\mathcal{G}} \, \Phi ^{\prime } \wedge e^{\wedge \Phi ^{\prime }}$, the zeros of ${\bf L} (e^{\wedge \Phi } )$ becomes equivalent to the zeros of $Q_{\mathcal{G}} \, \Phi ^{\prime }$ unless $\Phi ^{\prime } = 0$. 
By identifying these zeros with the on-shell condition of string field theory, we try to clarify the relation of two formulations. 
One can write ${\bf L} (e^{\wedge \Phi })=0$ for the on-shell condition in $L_{\infty }$-type formulation and $Q_{\mathcal{G}} \, \Psi _{\eta } = 0$ for that in WZW-like formulation, where the associated field $\Psi _{\eta }$ is a function of large-space string fields $V$. 
Therefore, the identification
\begin{align}
\Phi ^{\prime } \cong \Psi _{\eta }
\end{align}
provides the on-shell equivalence of $L_{\infty }$-type and WZW-like formulations. 
Furthermore, we show that at least up to quartic order, this identification $\Phi ^{\prime } \cong \Psi _{\eta }$ also provides the off-shell equivalence of two formulations for NS closed superstring field theory. 

\vspace{2mm}

This paper is organized as follows. 
In section 2, we briefly review a coalgebraic description of a cyclic $L_{\infty }$-algebra, which plays an important role in the action for (super-) string field theory. 
The technique of coalgebraic operations is a key tool in this paper. 
In section 3, first we introduce the concept of the path-ordered integral of multilinear maps, an iterated integral with directions. 
Then, we construct the NS string products ${\bf L}$ and the shifted BRST operator $Q_{\mathcal{G}}$ by appropriate similarity transformations of the BRST operator $Q$. 
This is the first result in this paper. 
As we will see in section 4, the similarity transformation naturally induces a morphism of two $L_{\infty }$-algebras. 
Using this morphism, one can derive the explicit form of the field redefinition $\Phi ^{\prime } \cong \Psi _{\eta }$ providing the on-shell equivalence of two formulations. 
Furthermore, we find that at least up to quartic order, $\Phi ^{\prime } \cong \Psi _{\eta }$ also provides the off-shell equivalence of two formulations. 
Then, a partial-gauge-fixing condition giving $L_{\infty }$-relations in WZW-like formulation naturally appears, which leads us to the idea that by choosing partial-gauge-fixing conditions of the WZW-like action, one can obtain any corresponding $L_{\infty }$-type actions. 
We end with conclusion and discussion. 

\vspace{2mm}

\section{Cyclic $L_\infty$-algebras}

In this section, we explain a cyclic $L_\infty$-algebra which are important tools to describe closed (super-) string field theory.
As a preparation for the following sections, 
we demonstrate it in closed bosonic string field theory.
See also \cite{Zwiebach:1992ie,Gaberdiel:1997ia,Lada:1992wc,Kajiura:2001ng,Getzler:2007} or mathematical manuscripts.

\niu{String products in closed bosonic SFT}

To begin with, we briefly review a bosonic closed string field theory constructed by Zwiebach \cite{Zwiebach:1992ie}.
The fundamental degree of freedom is a string field $\Psi$ which carries ghost number $2$ and is Grassmann even.
The action is given by
\begin{align}
S_\mathrm B=&\frac12 \langle \Psi, Q\Psi\rangle_{\mathrm B} +\sum_{n=2}^\infty \frac{\kappa^{n-1}}{(n+1)!}\langle \Psi, [\overbrace{\Psi,\Psi,...,\Psi}^{n}] \rangle_{\mathrm B},
\end{align}
where $\langle A ,B \rangle_{\mathrm B}$ is the BPZ inner product 
with $c_0^-=\frac12 (c_0 - \bar{c}_0)$ insertion, which satisfies
\begin{align}
\langle A,B\rangle_{\mathrm B}=(-)^{(A+1)(B+1)}\langle A,B\rangle_{\mathrm B}.\label{inner bos}
\end{align}
Owing to the anomaly in the conformal ghost sector, the inner product $\langle A,B\rangle_\mathrm B$ vanishes
unless the sum of the ghost number of $A$ and $B$ equals to $5$,
which imposes the condition that the $n$-string product carries ghost number $-2n+3$.
The kinetic term consists of the BRST operator $Q$ and so it gives the physical state condition in the first-quantization of strings.
The cubic and higher interaction vertices are given by the string products $[\Psi,...,\Psi]$, which satisfies the following algebraic properties:
\begin{align}
&\hspace{45pt}[B_{\sigma(1)},\dots,B_{\sigma(k)}]=(-)^\sigma [B_{1},\dots,B_k]\hspace{40pt}(\textrm{Graded commutativity}),\\[10pt]
&\hspace{10pt}0=
\sum_{i+j=n+1}{\sum_{\sigma}}'(-)^\sigma
[\: [B_{\sigma(1)},\dots,B_{\sigma(i)}],B_{\sigma(i+1)},\dots,B_{\sigma(n)} ]\label{L inf bos}\hspace{20pt}(L_\infty \textrm{ relations}),\\[2pt]
&\langle  B_1,[B_2,...,B_{n+1}]\rangle_{\mathrm B} = (-)^{B_1 +B_2+...+B_n }\langle [B_1,B_2,...,B_n],B_{n+1}\rangle_{\mathrm B}\label{cyc bos}\hspace{26pt}(\textrm{cyclicity}),
\end{align}
where 
$(-)^\sigma$ is the sign factor of the permutation from $\{B_1,...,B_n\}$ to $\{B_{\sigma(1)}, ... ,B_{\sigma(n)}\}$,
and we define $[B]= QB$,
and ${\sum_\sigma}'$ means the summation over all different $(i,n-i)$ splittings with $i\geq 1$.

The string products satisfying the $L_\infty$-relations play crucial roles, especially for the gauge invariance.
Utilizing their commutativity and cyclicity, the variation of the action can be taken. The equation of motion is given by
\begin{align}
\sum_{n=1}^\infty \frac{\kappa^{n-1}}{n!}[\overbrace{\Psi,...,\Psi}^{n}] =0.
\end{align}
From (\ref{L inf bos}),
one can see that the equation of motion belongs to the kernel of the operator $\sum_{m=0}^\infty \frac{\kappa^m}{m!}[\overbrace{\Psi,...,\Psi}^{m}, \;\cdot\;]$,
and this operator generate the gauge transformation of the field:
\begin{align}
\delta \Psi =\sum_{m=0}^\infty \frac{\kappa^m}{m!}[\overbrace{\Psi,...,\Psi}^{m}, \Lambda].
\end{align}
The gauge invariance follows from the cyclicity and the $L_\infty$-relations:
\begin{align}
\delta S
=\sum_{n=1}^\infty \frac{\kappa^{n-1}}{n!}\langle \delta\Psi, [\overbrace{\Psi,...,\Psi}^{n}]\rangle_{\mathrm B}
=\sum_{n=1}^\infty \frac{\kappa^{n-1}}{n!}\sum_{m=0}^\infty \frac{\kappa^m}{m!}
\langle \Lambda,[ [\overbrace{\Psi,...,\Psi}^{n}],\overbrace{\Psi,...,\Psi}^{m}]\rangle_{\mathrm B}
=0.
\end{align}
In addition to the gauge invariance,
the $L_\infty$-relations would be important also for the reproduction of the scattering amplitudes in the first-quantization of strings,
which are given by the single covering of the moduli spaces of punctured Riemann surfaces.
In bosonic string field theory, it is known that 
the string products defined to give the single covering of the moduli space
naturally satisfy the $L_\infty$-relations\cite{Sonoda:1989wa, Zwiebach:1992ie}.

\subsection{Coalgebras and multilinear maps}

\niu{Symmtrized tensor algebras as coalgebras}

Let $\cal C$ be a set.
When a coproduct $\Delta : \cal C \to C \otimes C$ is defined on $\cal C$ and it is coassociative
\begin{align}
(\Delta\otimes \1)\Delta = (\1\otimes \Delta)\Delta,
\end{align}
then $({\cal C},\Delta)$ is called a {\it coalgebra}.

In our case, $\cal C$ corresponds to the symmetrized tensor algebra $\cal S(H)$ of the $Z_2$-graded vector space $\cal H$.
In the language of closed string field theory, $\cal H$ is the state space for the string field,
and the $Z_2$-grading, called degree, equals to the Grassman parity.

The {\it symmetrized tensor product} $\wedge$ for elements of $\cal H$ is defined by
\begin{align}
\Phi_1\wedge\Phi_2
=\Phi_1\otimes\Phi_2
+(-)^{{\rm deg}(\Phi_1){\rm deg}(\Phi_2)}\Phi_2\otimes\Phi_1\:\:,\:\: \Phi_i\in\cal H.
\end{align}
This product satisfies the following properties:
\begin{align}
\Phi_1\wedge\Phi_2&=(-)^{{\rm deg}(\Phi_1){\rm deg}(\Phi_2)}\Phi_2\wedge\Phi_1,\\
(\Phi_1\wedge\Phi_2)\wedge\Phi_3&=\Phi_1\wedge(\Phi_2\wedge\Phi_3),\\
\Phi_1\wedge\Phi_2\wedge...\wedge \Phi_n
&=\sum_{\sigma}(-)^{\sigma}\Phi_{\sigma(1)}\otimes\Phi_{\sigma(2)}\otimes...\otimes \Phi_{\sigma(n)}.
\end{align}

We can construct a {\it symmetrized tensor algebra} $\cal S(H)$ by
\begin{align}
\mathcal {S(H)}=\mathcal H^{\wedge 0}\oplus\mathcal H^{\wedge 1}\oplus\mathcal H^{\wedge 2}\oplus\cdots.
\end{align}
We can define a coassociative coprduct $\Delta:{\cal S(H)}\to{\cal S(H)} \wedge{\cal S(H)}$ and a set $({\cal S(H)},\Delta)$ gives coalgebra. 
The action of $\Delta$ on $\Phi_1\wedge...\wedge \Phi_{n}\in {\cal H}^{\wedge n}$ is given by
\begin{align}
\Delta(\Phi_1\wedge...\wedge \Phi_{n})
=\sum_{k=0}^{n}{\sum_\sigma}'(-)^{\sigma}
(\Phi_{\sigma(1)}\wedge...\wedge \Phi_{\sigma(k)})
\wedge
(\Phi_{\sigma(k+1)}\wedge...\wedge \Phi_{\sigma(n)}).
\end{align}

\niu{Multi-linear maps}

From a multilinear map $b_n:\mathcal H^{n}\to \mathcal H$ which is graded symmetric upon the interchange of the arguments,
a map $b_n:\mathcal H^{\wedge n}\to \mathcal H$ is naturally defined by
\begin{align}
b_n(\Phi_1\wedge\Phi_2\wedge...\wedge \Phi_n)=b_n(\Phi_1,\Phi_2,...,\Phi_n).
\end{align}
The symmetric tensor product of two multilinear maps $A:\mathcal H^{\wedge k}\to \mathcal H^{\wedge l}$ and $B:\mathcal H^{\wedge m}\to \mathcal H^{\wedge n}$,
$A\wedge B:\mathcal H^{\wedge k+m}\to \mathcal H^{\wedge l+n}$,
can also be defined naturally by
\begin{align}
A\wedge B(\Phi_1\wedge...\wedge \Phi_{k+m})
={\sum_\sigma}'(-)^{\sigma}
A(\Phi_{\sigma(1)}\wedge...\wedge \Phi_{\sigma(k)})
\wedge
B(\Phi_{\sigma(k+1)}\wedge...\wedge \Phi_{\sigma(k+m)}).
\end{align}

The identity operator on ${\cal H}^{\wedge n}$ is defined by 
\begin{align}
\mathbb I_n=\frac1{n!}\mathbb I\wedge\mathbb I\wedge...\wedge\mathbb I=
\mathbb I\otimes\mathbb I\otimes...\otimes\mathbb I.
\end{align}
Note that we need the coefficient $\frac 1{n!}$.

Multilinear maps with degree $1$ and $0$ naturally induce the maps from $\cal S(H)$ to $\cal S(H)$.
They are called a coderivation and a cohomomorphism respectively, and are the main focus of the rest of this subsection.

\niu{Multi-linear maps as a coderivation}

A linear operator $m:{\cal C}\to {\cal C}$ which raise the degree one is called {\it coderivation} if it satisfies
\begin{align}
\Delta m = (m\otimes \1)\Delta +(\1\otimes m)\Delta.
\end{align}

From a map $b_n:{\cal H}^{\wedge n}\to\cal H$ which carries the degree one, 
the coderivation ${\bf b}_n:\mathcal {S(H)}\to\mathcal {S(H)}$ is naturally defined by
\begin{align}
{\bf b}_n\Phi =(b_n\wedge\mathbb I_{N-n})\Phi\:\:,\:\:\Phi \in {\cal H}^{\wedge N\geq n}\subset \mathcal {S(H)},
\end{align}
and ${\bf b}_n$ vanishes when acting on $ {\cal H}^{\wedge N\leq n}$.
We will call $\mathbf b_n$ a {\it $n$-coderivation }.

The explicit action of the one-coderivation ${\bf b}_1$ is given by
\begin{align}
\begin{array}{rccl}
{\bf b}_1 : & 1 & \to &0 \\
&\Phi_1&\to &b_1(\Phi_1)\\
&\Phi_1\wedge\Phi_2&\to &b_1(\Phi_1)\wedge \Phi_2 +(-)^{{\rm deg}(\Phi_1){\rm deg}(b_1)}\Phi_1\wedge b_1(\Phi_2).
\end{array}
\end{align}

We can also define the zero-coderivation ${\bf b}_0:\cal S(H)\to S(H)$ which is derived from a map $b_0:\mathcal H^{0}\to \mathcal H$ by
\begin{align}
{\bf b}_0 \Phi =(b_0\wedge\mathbb I_{N})\Phi =b_0\wedge\Phi\:\:,\:\:\Phi \in {\cal H}^{\wedge N}\subset \mathcal {S(H)}.
\end{align}
Its explicit action is as follows:
\begin{align}
\begin{array}{rccl}
{\bf b}_0:&1&\to&b_0\\
&\Phi_1&\to&b_0\wedge \Phi_1\\
&\Phi_1\wedge\Phi_2&\to &b_0\wedge\Phi_1\wedge \Phi_2 .
\end{array}
\end{align}
It is useful in the field redefinition and the gauge transformation.

Given two coderivations ${\bf b}_n$ and ${\bf c}_m$ which are derived from $b_n:{\cal H}^{\wedge n}\to \cal H$ and $c_m:{\cal H}^{\wedge m}\to \cal H$ respectively,
the graded commutator $[\![{\bf b}_n ,{\bf c}_m]\!]$ becomes the coderivation derived from 
the map $[\![b_n,c_m]\!]:{\cal H}^{\wedge n+m-1}\to \cal H$ which is defined by
\begin{align}
[\![b_n,c_m]\!]&=b_n(c_m\wedge\mathbb I_{n-1})-(-)^{{\rm deg}(b_n){\rm deg}(c_m)}c_m(b_n\wedge\mathbb I_{m-1}).
\end{align}

\niu{Multilinear maps as a cohomomorphism}

Given two coalgebras $C,C'$, a {\it cohomomorphism} ${\sf f}: C \to C'$ is a map of degree zero satisfying
\begin{align}
\Delta {\sf f} = ({\sf f}\otimes {\sf f})\Delta.\label{cohom}
\end{align}

A set of degree zero multilinear maps $ \{ \mathsf f_n : \mathcal H^{\wedge n}\to \mathcal H'\}_{n=0}^{\infty}$ 
naturally induces a cohomorphism $\mathsf f : \mathcal{S(H)}\to \mathcal{S(H')}$, which we denote as $\mathsf f= \{ \mathsf f_n\}_{n=0}^{\infty}$.
Its action on $\Phi _{1}\wedge \dots \wedge \Phi _{n} \in \mathcal H^{\wedge n}\subset  \mathcal{S(H)}$ is defined by
\begin{align}
\mathsf f (\Phi _{1}\wedge \dots \wedge \Phi _{n})=
\sum_{i \leq n} \sum_{k_{1} < \dots < k_{i}} 
&e^{\wedge \mathsf f_0}\wedge{\sf f}_{k_{1}} ( \Phi _{1} , \dots , \Phi _{k_{1}}) \wedge {\sf f}_{k_{2}-k_{1}} ( \Phi _{k_{1} +1} , \dots , \Phi _{k_{2}} ) \wedge \no
&\hspace{15pt}\dots \wedge  {\sf f}_{k_{i} -k_{i-1}} ( \Phi _{k_{i-1}+1} , \dots , \Phi _{n} ).
\end{align}

Its explicit actions are given as follows:
\begin{align}
\begin{array}{rccl}
\mathsf f :&1&\to& e^{\wedge \mathsf f_0}\\
&\Phi&\to&e^{\wedge \mathsf f_0}\wedge \mathsf f_1(\Phi) \\
&\Phi_1\wedge\Phi_2&\to &e^{\wedge \mathsf f_0}\wedge \mathsf f_1(\Phi_1)\wedge  \mathsf f_1(\Phi_2) 
+e^{\wedge \mathsf f_0}\wedge \mathsf f_2(\Phi_1\wedge\Phi_2) .
\end{array}
\end{align}

\subsection{Cyclic $L_\infty$-algebra}

\niu{Cyclic $L_\infty$-algebra $(\mathcal H, \mathbf L, \omega)$}

Let $\cal H$ be a graded vector space and $\cal S(H)$ be its symmetrized tensor algebra.
A {\it weak $L_\infty$-algebra} $(\mathcal{H},{\bf L})$ is a coalgebra $\cal S(H)$ with a coderivation ${\bf L} ={\bf L}_0+{\bf L}_1+{\bf L_2}+...$ satisfying
\begin{align}
({\bf L})^2=0.\label{L^2=0}
\end{align}
We denote the collection of the multilinear maps $\{L_k\}_{k\geq 0}$ also by $\bf L$.
In particular,  if ${\bf L}_0=0$, $(\mathcal{H},{\bf L})$ is called an {\it $L_\infty$-algebra}.

In the case of an $L_\infty$-algebra, the part of (\ref{L^2=0}) that correspond to an $n$-fold multilinear map $\mathcal H^{\wedge n}\to\mathcal H$ is given by
\begin{align}
{\bf L}_n\cdot{\bf L}_1+{\bf L}_{n-1}\cdot{\bf L}_2+\cdots+{\bf L}_2\cdot{\bf L}_{n-1}+{\bf L}_1\cdot{\bf L}_n=0.
\end{align}
We can act it on $B_1\wedge B_2\wedge ... \wedge B_n \in\mathcal H^{\wedge n}$ 
to get the $L_\infty$ relations for the multilinear maps $\{L_k\}$:
\begin{align}
0=\sum_{i+j=n+1}{\sum_{\sigma}}'(-)^\sigma
L_j( L_i(B_{\sigma(1)},\dots,B_{\sigma(i)}),B_{\sigma(i+1)},\dots,B_{\sigma(n)}).
\end{align}

We can define an inner product $\langle \cdot ,\cdot\rangle:\mathcal H^{\otimes 2}\to \mathbb C$ satisfying the same property as (\ref{inner bos})
using the graded symplectic form
$\langle\omega|:\mathcal H^{\otimes 2}\to \mathbb C$:
\begin{align}
\langle A,B\rangle
=(-)^A\langle\omega|A\otimes B.
\end{align}
Given the operator $\mathcal O_n$, we can define its BPZ-conjugation $O^\dagger_n$ as follows:
\begin{align}
\langle\omega|\mathbb I \otimes \mathcal O_n =
\langle\omega| \mathcal O^\dagger_n \otimes\mathbb I.
\end{align}

A set $({\cal S(H)},{\bf L},\omega)$ is called {\it cyclic $L_\infty$-algebra}
if each $L_n$ is BPZ-odd:
\begin{align}
L_n^\dagger = -L_n.
\end{align}

\niu{Projector and group-like element}

We can naturally define a {\it projector} ${\boldsymbol \pi}: {\cal S(H)}\to {\cal H}$
whose action on $\Phi\in\cal S(H)$ is given by
\begin{align}
{\boldsymbol \pi} \Phi = \Phi_1, 
\Phi =\sum_{n=1}^\infty \Phi _{1}\wedge \dots \wedge \Phi _{n} \in \mathcal{S(H)}.\label{pi}
\end{align}
Note that $\boldsymbol \pi$ acts trivially on $\cal H$ 
and commutes with one-coderivations.

Let ${\cal H}_0$ be the degree zero part of $\cal H$.
The following exponential map of $\Phi\in {\cal H}_0$ 
\begin{align}
e^{\wedge \Phi} = {\bf 1} +\Phi + \frac12 \Phi\wedge\Phi +\frac1{3!}\Phi\wedge\Phi\wedge\Phi +\cdots.
\end{align}
is called a {\it group-like element}. It satisfies
\begin{align}
\Delta e^{\wedge\Phi} =e^{\wedge\Phi}\wedge e^{\wedge\Phi},
\end{align}
 
The action of a coderivation on a group-like element is given by
\begin{align}
{\bf b}_n (e^{\wedge\Phi}) = \frac1{n!} b_n(\Phi^{\wedge n})\wedge e^{\wedge\Phi}.
\end{align}
Where we promise $0!=1$.
Note that we can not distinguish a one-coderivation $\mathbf b_1$ derived from a linear map $b_1:\mathcal H \to \mathcal H, \Phi \mapsto b_1(\Phi)$
and a zero-coderivation $\mathbf b_0$ derived from $b_0:{\cal H}^{\wedge 0} \to \mathcal H, 1\mapsto b_0=b_1(\Phi)$ 
when acting on group-like element:
\begin{align}
{\bf b}_0 (e^{\wedge\Phi})=b_0 \wedge(e^{\wedge\Phi})=b_1(\Phi) \wedge(e^{\wedge\Phi})={\bf b}_1 (e^{\wedge\Phi}).
\end{align}

One of the important property of a cohomomorphism is its action on the group-like element:
\begin{align}
\Delta \mathsf f (e^{\wedge \Phi}) =(\mathsf f \otimes \mathsf f)\Delta e^{\wedge \Phi}=(\mathsf f\otimes \mathsf f)e^{\wedge \Phi}\wedge e^{\wedge \Phi}
=\mathsf f(e^{\wedge \Phi})\wedge \mathsf f(e^{\wedge \Phi}).
\end{align}
We can see that the cohomomorphisms preserves the group-like element : $\mathsf f (e^{\wedge \Phi})=e^{\wedge \Phi'}$.

Utilizing the projector and the group-like element, 
the {\it Maurer-Cartan element} for an $L_\infty$-algebra $(\mathcal H, \mathbf L)$ is given by
\begin{align}
\mathcal F_\Phi
:=\boldsymbol \pi{\bf L} (e^{\wedge \Phi}) 
= \mathbf L_1(\Phi) + \frac12 \mathbf L_2(\Phi\wedge\Phi) +\frac1{3!}\mathbf L_3(\Phi\wedge\Phi\wedge\Phi) +\cdots.
\end{align}
The {\it Maurer-Cartan equation} for an $L_\infty$-algebra $(\mathcal H, \mathbf L)$ is given by
$\mathcal F_\Phi =0$,
which correspond to the on-shell condition in string field theory based on the $L_\infty$-algebra$(\mathcal H, \mathbf L)$.

\niu{Zwiebach's closed bosonic string field theory in coalgebraic representation}

To conclude this section, let us describe closed bosonic string field theory in the coalgebraic representation.
String products in closed bosonic string field theory
can be represented by a set of multilinear maps $L^\mathrm{B}_n:{\mathcal H_\mathrm{B}}^{\wedge n}\to \mathcal H_\mathrm{B}$:
\begin{align}
[\Psi_1,\Psi_2,...,\Psi_n]=L^\mathrm{B}_n(\Psi_1\wedge\Psi_2\wedge...\wedge \Psi_n),
\end{align}
and the set of $\{L^\mathrm{B}_n\}$ naturally define a set of coderivations $\{ \mathbf L^\mathrm{B}_n\}$.
Because of the $L_\infty$ relation (\ref{L inf bos}) for the original products, $\mathbf L^\mathrm{B} = \sum_{n=1}^\infty \mathbf L^\mathrm{B}_n$ is nilpotent
\begin{align}
({\bf L^\mathrm{B}})^2=0.
\end{align}
The cyclicity of original string products (\ref{cyc bos}) corresponds to ${L^\mathrm{B}_n}^\dagger = -L^\mathrm{B}_n$.
Therefore the algebraic properties of the string products is encoded to the fact that 
 $(\mathcal H_\mathrm{B},\mathbf L^\mathrm{B},\omega_\mathrm{B})$ defines a cyclic $L_\infty$ algebra.

We can transform the action into the form respecting its $L_\infty$-algebra.
Let $t$ be a real parameter $t \in [ 0 , 1 ]$. 
We introduce a $t$-parametrized string field $\Psi (t)$ satisfying $\Psi ( 0 ) = 0$ and $\Psi (1) = \Psi $,
which is a path connecting $0$ and the string field $\Psi $ in the space of string fields. 
Using this $\Psi(t)$, 
\begin{align}
S_\mathrm B
&=\frac12 \langle \Psi,Q\Psi\rangle +\sum_{n=2}^\infty \frac{\kappa^{n-1}}{(n+1)!}\langle \Psi, L^\mathrm{B}_n(\overbrace{\Psi,\Psi,...,\Psi}^{n}) \rangle_{\mathrm B}\no
&=\sum_{n=1}^\infty \frac{\kappa^{n-1}}{(n+1)!}\langle \Psi, L^\mathrm{B}_n(\overbrace{\Psi,\Psi,...,\Psi}^{n}) \rangle_{\mathrm B}\no
&=\int_0^1 dt \sum_{n=1}^\infty \frac{\kappa^{n-1}}{(n+1)!} \partial_t \langle \Psi(t),L^\mathrm{B}_n(\overbrace{\Psi(t),\Psi(t),...,\Psi(t)}^{n}) \rangle_{\mathrm B}\no
&=\int_0^1 dt \sum_{n=1}^\infty \frac{\kappa^{n-1}}{n!}\langle \partial_t\Psi(t),L^\mathrm{B}_n(\overbrace{\Psi(t),\Psi(t),...,\Psi(t)}^{n}) \rangle_{\mathrm B}\no
&=\int_0^1 dt \langle \partial_t\Psi(t),\mathcal F_{\Psi(t)} \rangle_{\mathrm B}.
\end{align}
In the fourth line we act $\partial_t$ and use the cyclicity of the string products to move $\partial_t\Psi(t)$ to the first slot of the inner product.
In the last line we represent the second slot of the inner product by the Maurer-Cartan element $\mathcal F_\Psi$ of the $L_\infty$-algebra $(\mathcal H_\mathrm{B},\mathbf L^\mathrm{B},\omega_\mathrm{B})$
 which is given by
\begin{align}
\mathcal F_\Psi
&=\sum_{n=1}^\infty \frac{\kappa^{n-1}}{n!}L^\mathrm{B}_n(\overbrace{\Psi,\Psi,...,\Psi}^{n}).
\end{align}
Utilizing the cyclicity, the variation of the action becomes $\delta S=\langle \delta\Psi, \mathcal F_\Psi \rangle_{\mathrm B}$
and the equation of motion is given by the Maurer-Cartan equation $\mathcal F_\Psi =0$.

The action can be represented using the coderivations and the group-like element as follows:
\begin{align}
S_\mathrm B
&=\int_0^1 dt \langle \partial_t\Psi(t),\boldsymbol \pi\mathbf L^\mathrm{B}(e^{\wedge\Psi(t)}) \rangle_{\mathrm B}\no
&=\int_0^1 dt \langle \boldsymbol \pi \boldsymbol\partial_t (e^{\wedge\Psi(t)}),\boldsymbol \pi\mathbf L^\mathrm{B}(e^{\wedge\Psi(t)}) \rangle_{\mathrm B},\label{action bos}
\end{align}
where we denote the one-coderivation derived from $\partial_t$ as $\boldsymbol\partial_t$.
Note that the group-like element we use in the first slot of the inner product of (\ref{action bos}) is just a representational convention.
However, this representation is the form respecting the $L_\infty$-algebra of the theory, 
and is very useful when we compare two actions.
We will see in the appendix A that WZW-like and $L_\infty$-type actions can be represented in this form.

In coalgebraic representation, the equation of motion is written as 
\begin{align}
\boldsymbol \pi \mathbf L^\mathrm{B}(e^{\wedge\Psi}) =0.
\end{align}
The equation of motion belongs to the kernel of the operator generating the gauge transformation.
In the present case, since $(\mathbf L^\mathrm B)^2=0$, the equation of motion belongs to the kernel of 
$\boldsymbol \pi\mathbf L^\mathrm{B}(e^{\wedge\Psi} \wedge \; \cdot \;)$:
\begin{align}
\boldsymbol\pi\mathbf L^\mathrm{B}\big(e^{\wedge\Psi} \wedge \boldsymbol \pi \mathbf L^\mathrm{B}(e^{\wedge\Psi}) \big)
=\boldsymbol\pi\mathbf L^\mathrm{B}\mathbf L^\mathrm{B}(e^{\wedge\Psi}) =0.
\end{align}
It means that the gauge transformation of the field is generated by
$\boldsymbol \pi\mathbf L^\mathrm{B}(e^{\wedge\Psi} \wedge \; \cdot \;)$,
and we can confirm it actuary is:
\begin{align}
\delta \Psi =\sum_{m=0}^\infty \frac{\kappa^m}{m!}[\overbrace{\Psi,...,\Psi}^{m}, \Lambda]
=\boldsymbol \pi\mathbf L^\mathrm{B}(e^{\wedge\Psi} \wedge \Lambda).
\end{align}

\section{Similarity transformations}

We focus on
two popular and successful formulations of NS superstring field theories:
the $L_\infty$-type formulation which is based on the small Hilbert space,
and the WZW-like formulation which is based on the large Hilbert space.
In both formulations, nonlinear nilpotent operators play crucial roles.
The first one $\bf L$ is the NS superstring products in the $L_\infty$-type formulation.
The second one $Q_\mathcal{G}$ is the BRST operator shifted by the bosonic pure gauge string field $\mathcal G$ in the WZW-like formulation.
In this section, we show that these nilpotent operators $\bf L$ and $Q_\mathcal{G}$ can be given by the similarity transformations from the BRST operator $Q$.

\niu{Path-ordered exponential}

The similarity transformations which we introduce in this section are given by the path-ordered exponential.
It is defined by the following iterated integral
\begin{align}
\mathcal A{\scriptstyle[\tau]}
&=\overset{\rightarrow}{\mathcal P} \exp \left(\int_0^\tau d\tau' \mathcal O{\scriptstyle{\scriptstyle[\tau']}}\right)\no
&=\1 +
\left(\int_0^{\tau} d\tau_1 {\mathcal O}{\scriptstyle[\tau_1]}\right)
+ \sum_{n=2}^\infty
\left(\int_0^{\tau} d\tau_1 {\mathcal O}{\scriptstyle[\tau_1]}\right)
\left(\int_0^{\tau_1} d\tau_2 {\mathcal O}{\scriptstyle[\tau_2]}\right)
\cdots
\left(\int_0^{\tau_{n-1}} d\tau_n {\mathcal O}{\scriptstyle[\tau_n]}\right).
\end{align}
The $\rightarrow$ over $\mathcal P$ denote the order of integrations.
This $\mathcal A$ satisfies the following differential equation
\begin{align}
\partial_\tau \mathcal A{\scriptstyle[\tau]} = \mathcal O{\scriptstyle[\tau]} \cdot  \mathcal A{\scriptstyle[\tau]}
\end{align}
and the initial condition 
$\mathcal A{\scriptstyle[0]}=\1$.
Note that we use the character $\tau$ to represent the parameter used in the iterated integral,
and that the dependence on $\tau$ is denoted by ${\scriptstyle[\:\:]}$ in order to distinguish it from the parameter in string field like $\Phi(t)$.
We will omit ${\scriptstyle[1]}$ for notational simplicity: ${\cal A} ={\cal A}\scriptstyle[1]$.

For the operator $\mathcal O$ which can be expanded in powers of $\tau$ as follows:
\begin{align}
\mathcal O{\scriptstyle [\tau]} = \sum_{k=0}^\infty \tau^k \mathcal O_{k+2},
\end{align}
the lower order terms in $\tau$ of
$\mathcal A{\scriptstyle [\tau]}=\overset{\rightarrow}{\mathcal P} \exp \left(\int_0^\tau d\tau' \mathcal O{\scriptstyle{\scriptstyle[\tau']}}\right)$
are given by
\begin{align}
\mathcal A {\scriptstyle [\tau]}=\1 + \tau \mathcal O_2 +\frac{\tau^2}{2}(\mathcal O_3 + \mathcal O_2 \mathcal O_2)
+\frac{\tau^3}{3!}(2 \mathcal O_4 + 2 \mathcal O_3 \mathcal O_2 +  \mathcal O_2 \mathcal O_3 + \mathcal O_2 \mathcal O_2 \mathcal O_2)
+\cdots.
\end{align}
Note that if $\mathcal O$ is independent of $\tau$, it becomes an usual exponential.

The inverse of $\mathcal A$ can be defined by $\mathcal O \to - \mathcal O$ with the reversal of the order of the integrations:
\begin{align}
\mathcal A^{-1}{\scriptstyle[\tau]}
&
=\overset{\leftarrow}{\mathcal P} \exp \left(-\int_0^\tau d\tau' \mathcal O{\scriptstyle[\tau']}\right)\no
&=\1 
-\left(\int_0^{\tau} d\tau_1 {\mathcal O}{\scriptstyle[\tau_1]}\right)
+ \sum_{n=2}^\infty(-)^n
\left(\int_0^{\tau_{n-1}} d\tau_n {\mathcal O}{\scriptstyle[\tau_n]}\right)
\cdots
\left(\int_0^{\tau_1} d\tau_2 {\mathcal O}{\scriptstyle[\tau_2]}\right)
\left(\int_0^{\tau} d\tau_1 {\mathcal O}{\scriptstyle[\tau_1]}\right).\label{A dag}
\end{align}
The integration in the equation (\ref{A dag}) is defined to be performed from the right to the left.
It can be represented in the usual representation as follows:
\begin{align}
\mathcal A^{-1}{\scriptstyle[\tau]}=\1 -
\int_0^{\tau} d\tau_1 {\mathcal O}{\scriptstyle[\tau_1]}
+ \sum_{n=2}^\infty (-)^n
\int_0^{\tau} d\tau_1\int_0^{\tau_1} d\tau_2\cdots \int_0^{\tau_{n-1}} d\tau_n 
{\mathcal O}{\scriptstyle[\tau_n]}\cdots{\mathcal O}{\scriptstyle[\tau_2]} {\mathcal O}{\scriptstyle[\tau_1]}.
\end{align}
The important property of $\mathcal A^{-1}$ is that it is the solution to the following equation
\begin{align}
\partial_\tau \mathcal A^{-1}{\scriptstyle[\tau]} = -\mathcal A^{-1}{\scriptstyle[\tau]}\cdot\mathcal O{\scriptstyle[\tau]}
\end{align}
with the initial condition $A^{-1}{\scriptstyle[0]}=\1$.

We can show that $\mathcal A^{-1}$ defined by (\ref{A dag}) gives actually the inverse of $\mathcal A$ 
by solving the differential equations for 
 $\mathcal A^{-1}{\scriptstyle[\tau]} \mathcal A{\scriptstyle[\tau]}-\1$:
\begin{align}
&\partial_\tau \Big(\mathcal A^{-1}{\scriptstyle[\tau]} \mathcal A{\scriptstyle[\tau]}-\1\Big)
=(-\mathcal A^{-1}{\scriptstyle[\tau]}\mathcal O{\scriptstyle[\tau]}\mathcal A){\scriptstyle[\tau]})+\mathcal A^{-1}{\scriptstyle[\tau]}(\mathcal O{\scriptstyle[\tau]}\mathcal A{\scriptstyle[\tau]})
=0.\label{a inv de 2}
\end{align}
Since the initial condition is given by $\mathcal A^{-1}{\scriptstyle[0]}\mathcal A{\scriptstyle[0]}-\1=0$,
the solution of the differential equation is $\mathcal A^{-1}{\scriptstyle[\tau]}\mathcal A{\scriptstyle[\tau]}-\1=0$,
which leads to 
\begin{align}
\mathcal A^{-1}{\scriptstyle[\tau]}\mathcal A{\scriptstyle[\tau]}=\1\label{a dag a}.
\end{align}
Acting $\mathcal A{\scriptstyle[\tau]}$ from the left and $\mathcal A^{-1}{\scriptstyle[\tau]}$ from the right on (\ref{a dag a}),
$\mathcal A{\scriptstyle[\tau]} \mathcal A^{-1}{\scriptstyle[\tau]}=\1$ can also be obtained.

\niu{Simirality transformation}

Given a linear operator $\hat q $, we can define its similarity tramsformation by $\mathcal A$ as
\begin{align}
\hat q ^\mathcal A= \mathcal A \, \hat q \, \mathcal A^{-1}.\label{st}
\end{align}
It has two important properties.
The first one is that it gives the solution of the following differential equation
\begin{align}
\partial_\tau \hat q^\mathcal A = [\![ \mathcal O, \hat q^\mathcal A\,]\!] \label{general DE}
\end{align}
with the initial condition $\hat q^\mathcal A {\scriptstyle [\tau =0]}=\hat q$.
The second one is as follows:
\begin{align}
( \hat q ^\mathcal A)^n=(\hat q^n) ^\mathcal A.
\end{align}
In particular, if $\hat q $ is nilpotent, $\hat q ^\mathcal A$ is also nilpotent.

In string field theory,  
if the operator $\mathcal O$ in the exponent is BPZ-odd: $\mathcal O^\dagger =-\mathcal O$,
the BPZ-conjugation of the path-ordered integral gives its inverse: $\mathcal A^{-1}=\mathcal A^{\dagger}$.
(Recall that the BPZ-conjugation $\dagger$ includes the inversion of the order of operators.)
In what follows, we see that the key nilpotent operators $\bf L$ in the $L_\infty$-type formulation and $Q_\mathcal G$ in the WZW-like formulation
satisfy the differential equations of the type of (\ref{general DE}),
and can be obtained by the similarity transformations of the BRST operator $Q$.

\subsection{Superstring product in the small Hilbert space}

To construct a consistent superstring field theory of NS sector in the small Hilbert space,
the insertion of the picture changing operator $X$ seems to be necessary.
In the works \cite{Erler:2014eba}, 
the NS superstring products $\bf L$
satisfying the $L_\infty$ relations are constructed in the systematic way
from the bosonic string products $\mathbf L^\mathrm{B}$ 
and the zero-modes $X,\xi$ of the picture changing operator $X(z)$ and the fermionized superconformal ghost $\xi(z)$.
In this subsection, we show that the NS superstring products $\bf L$ can be obtained by the similarity transformation of the BRST operator $Q$.

In the small space formulation, the fundamental degree of freedom is the string field $\Phi$ carrying ghost number $2$ and picture number $-1$,
belonging to the small Hilbert space $\mathcal H_\mathrm{small}$: $\eta\Phi=0$.
The action is written using the NS string products $\mathbf L = \{L_k\}_{k\geq 1}$:
\begin{align}
S_{\scriptscriptstyle\rm EKS}
&=\frac12\langle \xi\Phi, Q\Phi \rangle+\sum_{n=2}^\infty \frac{\kappa^{n-1}}{(n+1)!}\langle \xi\Phi, L_n(\overbrace{\Phi,\Phi,...,\Phi}^{n}) \rangle\no
&=\sum_{n=1}^\infty \frac{\kappa^{n-1}}{(n+1)!}\langle \xi\Phi, L_n(\overbrace{\Phi,\Phi,...,\Phi}^{n}) \rangle,
\end{align}
where $L_1=Q$ and the inner product is the $c_0^-$-inserted BPZ inner product satisfying
\begin{align}
\langle A,B\rangle =(-)^{(A+1)(B+1)}\langle B,A\rangle.
\end{align}
Owing to the anomaly in the superconformal ghost sector, the inner product $\langle A,B\rangle$ vanishes
unless the sum of the ghost number of $A$ and $B$ equals to $4$ and the sum of the picture number of $A$ and $B$ equals to $-1$.
These anomalies impose the condition that the NS string product $L_n$ carries ghost number $-2n+3$ and picture number $n-1$.
Besides, the NS string products $\bf L$ satisfy the following three properties:
\begin{eqnarray}
&{\bf L}^2=0 &(L_\infty \textrm{ relation}),\label{L infty}\label{eks l inf}\\ 
&{\bf L}^\dagger =-{\bf L}&(\textrm{cyclicity}),\label{eks cyc}\\
&\ld {\bf L},\eta\rd=0&(\eta\textrm{-derivation}).\label{eks eta der}
\end{eqnarray}
The theory is characterized by the cyclic $L_\infty$-algebra $(\mathcal H_\mathrm{small}, \mathbf L, \omega)$.

As in the case of closed bosoninc string field theory, the action can be transformed into the form respecting the $L_\infty$-algebra.
Let $t$ be a real parameter $t \in [ 0 , 1 ]$. 
We introduce a $t$-parametrized string field $\Phi (t)$ satisfying $\Phi ( 0 ) = 0$ and $\Phi (1) = \Phi $, which is a path connecting $0$ and the string field $\Phi $ in the space of string fields. 
Using this $\Phi (t)$, we can rewrite the $L_{\infty }$-type action as follows. 
\begin{align}
S_{\scriptscriptstyle \rm EKS } & = \frac{1}{2} \langle \xi \Phi , \, Q \Phi \rangle +  \sum_{n=1}^{\infty } \frac{\kappa ^{n} }{(n+2)!} \langle \xi \Phi , \, L_{n+1} ( \overbrace{\Phi , \dots , \Phi }^{n+1} ) \rangle  
\no 
& = \int_{0}^{1} dt \, \frac{\partial }{\partial t} \Big( \sum_{n=0}^{\infty } \frac{\kappa ^{n} }{(n+2)!} \langle \xi \Phi (t) , L_{n+1} ( \overbrace{\Phi (t) , \dots , \Phi (t)}^{n+1} ) \rangle  \Big) 
\no 
& = \int_{0}^{1} dt \, \langle  \xi \partial _{t} \Phi (t) , \, \mathcal{F} _{\Phi (t)} \rangle , 
\end{align}
where $\mathcal{F}_{\Phi (t)}$ is the Maurer-Cartan element of the $L_{\infty }$-algebra 
\begin{align}
\mathcal{F}_{\Phi (t) } := Q \Phi (t) + \sum_{n=1}^{\infty } \frac{\kappa ^{n}}{(n+1)!} L_{n+1} ( \overbrace{\Phi (t) , \dots , \Phi (t)}^{n+1} )  , 
\end{align}
which gives the on-shell condition of $L_{\infty }$-type superstring field theory $\mathcal{F}_{\Phi }=0$. 
Note that since the variation of the action becomes $\delta S_{\scriptscriptstyle\rm EKS} = \langle \xi \delta \Phi , \mathcal{F}_{\Phi } \rangle $, the $t$-dependence is topological. 

In the coalgebraic representation, the action becomes   
\begin{align}
S_{\scriptscriptstyle \rm EKS } & = \int_{0}^{1} dt \, \langle {\boldsymbol \pi } ( {\boldsymbol \xi }_{t} \, e^{\wedge \Phi (t)} ) , {\boldsymbol \pi } \big( {\bf L}( e^{\wedge \Phi (t)} ) \big) \rangle ,
\end{align}
where $\boldsymbol{\xi}_t$ is a one-coderivation derived from $\partial_t\xi$.
Its explicit action is given by
\begin{align}
\begin{array}{rccc}
\boldsymbol\xi_t : & 1 & \to &0 \\
&\Phi&\to &\partial_t\xi\Phi\\
&\Phi_1\wedge\Phi_2&\to &(\partial_t\xi\Phi_1)\wedge \Phi_2 + \Phi_1\wedge(\partial_t\xi\Phi_2).
\end{array}\label{Xi t}
\end{align}
Note that $\partial_t$ and $\xi$ act on the same slot.
At this stage one can see that the $L_\infty$-type action consists of the $L_\infty$-algebra $(\Phi, \mathbf L)$.

In coalgebraic representation, the equation of motion can be written as
\begin{align}
\boldsymbol\pi{\bf L}( e^{\wedge \Phi (t)} ) =0.
\end{align}
The equation of motion belongs to the kernel of the operator generating the gauge transformation.
In the present case, since $\mathbf L^2=0$, the equation of motion belongs to the kernel of 
$\boldsymbol \pi\mathbf L(e^{\wedge\Phi} \wedge \;\cdot\;)$:
\begin{align}
\boldsymbol \pi\mathbf L\big(e^{\wedge\Phi} \wedge \boldsymbol\pi{\bf L}( e^{\wedge \Phi (t)} )\big)
=\boldsymbol \pi \mathbf L\mathbf L(e^{\wedge\Phi}) =0.
\end{align}
It means that the gauge transformation of fields is generated by $\mathbf L$:
\begin{align}
\delta \Phi =\boldsymbol \pi\mathbf L(e^{\wedge\Phi} \wedge \Lambda),
\end{align}
where $\Lambda$ is the gauge parameter carrying ghost number $1$ and picture number $-1$.
In addition, since $[\![\boldsymbol\eta,\mathbf L]\!]=0$, the equation of motion belongs also to the kernel of $\boldsymbol \eta$:
\begin{align}
\boldsymbol \eta \mathbf L(e^{\wedge\Phi}) =0.
\end{align}
However, since string fields belong to the kernel of $\eta$, 
it does not generate the gauge transformation of the fields in the small Hilbert space.
When we replace the string field $A$ in the small Hilbert space by $A = \eta B$ using the string field $B$ in the large Hilbert space
it corresponds to the gauge transformation of the large-space string field $B$, which does not change the small-space string field $A$.

\niu{$L_\infty$-products from the similarity transformation of $Q$}

Let us first consider the defining equation for the string products $\bf L$ satisfying $L_\infty$-relations.
Introducing a parameter $\tau$ as
${\bf L}{\scriptstyle[\tau]}=\sum_{n=1}^\infty \tau^{n-1} {\bf L}_{n}$,
and differentiating ${\bf L}{\scriptstyle[\tau]}^2$ by $\tau$:
\begin{align}
\partial_\tau \big({\bf L}{\scriptstyle[\tau]}^2\big)
= \big(\partial_\tau {\bf L}{\scriptstyle[\tau]}\big) {\bf L}{\scriptstyle[\tau]}
+{\bf L}{\scriptstyle[\tau]}\big(\partial_\tau {\bf L}{\scriptstyle[\tau]}\big)
=\Ld {\bf L}{\scriptstyle[\tau]},\big(\partial_\tau {\bf L}{\scriptstyle[\tau]}\big)\Rd,\label{DE1}
\end{align}
one can find that ${\bf L}{\scriptstyle[\tau]}^2=0$ hold if 
${\bf L}{\scriptstyle[0]}^2=0$
and $\partial_\tau{\bf L}{\scriptstyle[\tau]}$ commute with $\mathbf L{\scriptstyle[\tau]}$ under ${\bf L}{\scriptstyle[\tau]}^2=0$.
One natural choice is 
\begin{align}
\partial_\tau{\bf L}{\scriptstyle[\tau]}=\Ld {\bf L}{\scriptstyle[\tau]},{\bf \Xi}{\scriptstyle[\tau]} \Rd,\label{L infty DE}
\end{align}
where ${\bf \Xi}{\scriptstyle[\tau]}$ is a set of BPZ-odd {\it gauge products}:
\begin{align}
{\bf \Xi}{\scriptstyle[\tau]}=\sum_{n=2}^\infty \tau^{n-2}{\bf \Xi}_{n}.
\end{align}
So far, introducing the gauge products $\bf \Xi$,
we can write down the defining equation for the string products $\mathbf L{\scriptstyle[\tau]}$ satisfying $L_\infty$ relation ${\bf L}{\scriptstyle[\tau]}^2=0$.

In appendix $C$ of \cite{Erler:2013xta}, the solution for (\ref{L infty DE}) is obtained for the theory with only one- and two- bosonic string products.
The procedure there can be extended to the theories in which original bosonic products consist of three- and more- string products.
The path-ordered exponentials discussed in the beginning of this section play a curtail role.
Since (\ref{L infty DE}) is of the form of (\ref{general DE}), the solution is given by the similarity transformation of $Q$:
\begin{align}
{\bf L} &= {\bf G} {\bf Q} {\bf G}^\dagger,
\end{align}
where $\mathbf G$ is the path-ordered exponential of the gauge products $\bf  \Xi$:
\begin{align}
{\bf G}{\scriptstyle[\tau]}=&\overset{\rightarrow}{\mathcal P} \exp \left(- \int_0^\tau d\tau' {\bf \Xi}{\scriptstyle[\tau']}\right).
\label{G}
\end{align}
Since the gauge products ${\bf \Xi}{\scriptstyle[\tau]}$ are BPZ-odd, $\mathbf G^\dagger$ is given by
\begin{align}
{\bf G} ^\dagger{\scriptstyle[\tau]}
=\overset{\leftarrow}{\mathcal P} \exp \left(\int_0^\tau d\tau' {\bf \Xi}{\scriptstyle[\tau']}\right), 
\end{align}
and it satisfies 
\begin{align}
{\bf G}^\dagger{\scriptstyle[\tau]}&={\bf G}{\scriptstyle[\tau]}^{-1}.
\end{align}
Utilizing the properties of the path-ordered exponential, we can see that 
${\bf L} = {\bf G} {\bf Q} {\bf G}^\dagger$ is the solution for (\ref{L infty DE}) with ${\bf L}{\scriptstyle[\tau=0]}=\mathbf Q$:
\begin{align}
\partial_\tau {\bf L}{\scriptstyle[\tau]}
=-{\bf\Xi}{\scriptstyle[\tau]}{\bf G}{\scriptstyle[\tau]} {\bf Q}{\bf G}^\dagger{\scriptstyle[\tau]} +{\bf G}{\scriptstyle[\tau]}{\bf Q}{\bf G}^\dagger{\scriptstyle[\tau]}{\bf\Xi}{\scriptstyle[\tau]}
=\Ld{\bf L}{\scriptstyle[\tau]},{\bf\Xi}{\scriptstyle[\tau]}\Rd
\end{align}
and that ${\bf L}$ is nilpotent more directly:
\begin{align}
{\bf L}^2 
={\bf G} {\bf Q} {\bf G}^\dagger{\bf G} {\bf Q} {\bf G}^\dagger
={\bf G} {\bf Q} {\bf Q} {\bf G}^\dagger
=0,
\end{align}
and furthermore that this $\mathbf L$ is BPZ-odd:
\begin{align}
\mathbf L^\dagger 
=(\mathbf G \mathbf Q \mathbf G^\dagger )^\dagger
=-\mathbf G \mathbf Q \mathbf G^\dagger =-\mathbf L.
\end{align}
Thus, we obtain the string ploducts $\mathbf L$ satisfying the $L_\infty$-relation (\ref{eks l inf}) and the cyclicity (\ref{eks cyc}) by introducing the BPZ-odd gauge products $\bf \Xi$.

\niu{$L_\infty$-products in EKS theories}

The $\eta$-derivation property (\ref{eks eta der}) and the quantum numbers of $\bf L$ follows from the detail of the gauge products $\bf \Xi$.
The $n$-th gauge product $\Xi_n$ must carry ghost number $-2n+2$ and picture number $n-1$.
In the work \cite{Erler:2014eba}, the recursive construction of the BPZ-odd gauge products $\bf \Xi$
which lead to the $\eta$-derivation property of $\bf L$ (\ref{eks eta der}) is given.
Here we do not explain the detail of their construction and only show their explicit forms:
\begin{align}
\label{EKS gauge 2}
\Xi_2(\Phi,\Phi) =& \frac13\Big(\xi[\Phi,\Phi]-2[\xi \Phi,\Phi]\Big),\\
\label{EKS gauge 3}
\Xi_3(\Phi,\Phi,\Phi)=
&\frac18\xi X[\Phi,\Phi,\Phi]+\frac38\xi [X\Phi,\Phi,\Phi]-\frac38[X\xi \Phi,\Phi,\Phi]
-\frac38X[\xi \Phi,\Phi,\Phi]-\frac34[X\Phi,\xi\Phi,\Phi]\no
&+\frac12\xi[\Phi,\xi[\Phi,\Phi]]-\frac12[\xi\Phi,\xi[\Phi,\Phi]]-[\Phi,\xi[\xi\Phi,\Phi]].
\end{align}
The string product $\bf L$ can be constructed from this $\bf \Xi$, and the explicit forms of $\bf L_2$ and $\bf L_3$ are
\begin{align}
L_2(\Phi,\Phi)=&\frac13\Big(X[\Phi,\Phi]+2[X\Phi,\Phi]\Big),\\
L_3(\Phi,\Phi,\Phi)=
&\frac1{16}\Big(X^2[\Phi,\Phi,\Phi]+3[X^2\Phi,\Phi,\Phi]\Big)+\frac38\Big([\Phi,X\Phi,X\Phi]+X[X\Phi,\Phi,\Phi]\Big)\no
&-\frac13[\xi X[\Phi,\Phi],\Phi]+\frac1{12}\Big(\xi[X[\Phi,\Phi],\Phi]+2[X[\xi \Phi,\Phi],\Phi]+[X[\Phi,\Phi],\xi \Phi]\Big)\no
&+\frac{-5}{12}\Big(X[\xi[\Phi,\Phi],\Phi]+2[\xi[X\Phi,\Phi],\Phi]+[\xi[\Phi,\Phi],X\Phi]\Big)\no
&+\frac{1}{24}\Big([[\Phi,\xi\Phi],X\Phi]+[[\Phi,X\Phi],\xi \Phi]+\xi[[X\Phi,\Phi],\Phi]+X[[\xi \Phi,\Phi],\Phi]\Big)\no
&+\frac{3}{16}\Big(X[[\Phi,\Phi],\xi \Phi]+2[[\xi \Phi,X\Phi],\Phi]+\xi[[\Phi,\Phi],X\Phi]\no
&\hspace{48pt}+X\xi[[\Phi,\Phi],\Phi]+2[[\xi X\Phi,\Phi],\Phi]+[[\Phi,\Phi],\xi X\Phi]\Big).
\end{align}

\subsection{Superstring product in the large Hilbert space}

In this subsection, we explain the key structure in the WZW-like formulation
which is a successful formulation for superstring field theories in the large Hilbert space.
Although we take the notation assuming heterotic string field theory, it works also for open string and type II string.

To begin with, recall that in bosonic closed string field theory, we can define the BRST operator and the string products around the new background $A$:
\begin{align}
Q_A B= & QB+\sum_{n=1}^{\infty}\frac{\kappa^n}{n!}[A^n,B],\\
[B_1,B_2,...,B_m]_{A}
=&\sum_{n=0}^{\infty}\frac{\kappa^n}{n!}[A^n,B_1,B_2,...,B_m].
\end{align}
This shifted products satisfy the weak $L_\infty$-relation and if $A$ satisfies the equation of motion of bosonic theory, it satisfies the $L_\infty$-relation.

The key ingredient in WZW-like theories is the (bosonic) pure gauge string field $\mathcal G$ which is the solution for the equation of motion of {\it bosonic} string field theory.
It is obtained by the successive infinitesimal gauge transformation from $0$ along the gauge orbit parameterized by $\tau$.
The pure gauge string field $\mathcal G$ in the WZW-like theory is obtained
by replacing the gauge parameter in bosonic theory with the string field $V$ in WZW-like theory.
Note that $V$ is Grassman odd and carries ghost number $1$ and picture number $0$, same as the gauge parameter in bosonic theory.
By construction, $\mathcal G$ is defined by the following differential equation:
\begin{align}
\partial_\tau \mathcal G{\scriptstyle[\tau]} =  Q_{\mathcal G\scriptscriptstyle[\tau]}V,
\end{align}
where $Q_\mathcal G$ is the BRST operator shifted by the pure gauge string field $\mathcal G$, 
\begin{align}
Q_{\mathcal G}B=&QB+\sum_{n=1}^{\infty}\frac{\kappa^n}{n!}[{\mathcal G}^n,B].
\end{align}
The pure gauge string field $\mathcal G\scriptstyle[\tau]$ and the $\mathcal G$-shifted BRST operator $Q_{\mathcal G\scriptscriptstyle[\tau]}$ satisfy
\begin{align}
\mathcal G {\scriptstyle[0]}&=0,\\
Q_{\mathcal G \scriptscriptstyle[0]}&=Q,\\
\mathcal G{\scriptstyle [\tau]}&=\int^\tau_0d\tau' Q_{\mathcal G\scriptscriptstyle[\tau']}V.
\end{align}
Their explicit forms are given by
\begin{align}
\mathcal G{\scriptstyle [\tau]}=\:&\tau QV +\frac{\kappa\tau^2}{2}[V,QV]+\frac{\kappa^2\tau^3}{3!}\big([V,QV,QV]+[V,[V,QV]]\Big)+\cdots,\\
Q_{\mathcal G\scriptscriptstyle [\tau]}B
=\: & QB +\kappa\tau[QV,B]+\frac{\kappa^2\tau^2}{2}\Big([[V,QV],B]+[QV,QV,B]\Big)\no
& \! +\frac{\kappa^3\tau^3}{6}\Big([[V,QV,QV],B]+[[V,[V,QV]],B]+3 [QV,[V,QV],B]+[QV,QV,QV,B]\Big)\no
&\!+\cdots.
\end{align}

To constract the WZW-like action, 
it is convenient to introduce a $t$-parametrized string field $V (t)$ satisfying $V ( 0 ) = 0$ and $V (1) = V $.
Using this $V (t)$ and $\mathcal G$-shifted BRST operator $Q_{\mathcal G\scriptscriptstyle[\tau]}$,
the action can be written as follows:
\begin{align}
S_{\scriptscriptstyle{\rm WZW}}
=\int_0^1 dt \langle \Psi_{\partial_t}(t), Q_{\mathcal G(t)} \Psi_\eta (t)\rangle.
\end{align}
Here $\Psi_\mathbb X$ for $\mathbb X={\eta, \partial_t, \delta}$ is an associated string field, which is defined by the following differential equation
\begin{align}
\partial_\tau \Psi_\mathbb X{\scriptstyle [\tau]}=\mathbb XV + \kappa \big[V,\Psi_\mathbb X{\scriptstyle [\tau]}\big]_{\mathcal G\scriptscriptstyle [\tau]},
\label{associated field}
\end{align}
and it satisfies
\begin{align}
Q_{\mathcal G}\Psi_\mathbb X =(-)^\mathbb X \mathbb X\mathcal G.
\end{align}

The variation of the WZW-like action can be taken as
$\delta S_{\scriptscriptstyle{\rm WZW}}
=\langle \Psi_{\delta}, Q_{\mathcal G} \Psi_\eta \rangle$
and the on-shell condition is given by
\begin{align}
Q_{\mathcal G} \Psi_\eta=0.
\end{align}
Note that $t$-dependence is topological as in the case of the $L_\infty$-type action.
Since $Q_{\mathcal G}\Psi_\eta =-\eta\mathcal G$,
the on-shell condition $Q_\mathcal G \Psi_\eta$ belongs to the kernel of $Q_\mathcal G$ and $\eta$,
which generate the gauge transformation of the composite field in WZW-like action:
\begin{align}
\Psi_\delta = Q_{\mathcal G}\lambda + \eta \omega.
\end{align}
The gauge invariance follows from the nilpotency of $Q_{\mathcal G}$ and $\eta$.
Especially, the nilpotency of $Q_\mathcal G$ plays an important role:
the WZW-like theory is characterized by the $L_\infty$-algebra $(\mathcal H', Q_\mathcal G )$ where $\mathcal H'$ is the space for $\Psi_\eta$,
which we will explain in section 4.

\niu{$Q_\mathcal G$ from the similarity transformation of $Q$}

Let us first consider the defining equation for the $\mathcal G$-shifted BRST operator $Q_\mathcal G$.
Since $\tau$-dependence of $Q_\mathcal G$ comes from the pure gauge string field $\mathcal G$, 
differentiation of $Q_\mathcal G$ by $\tau$ gives
\begin{align}
\partial_\tau\big( Q_{\mathcal G\scriptscriptstyle [\tau]}B\big) 
&= \sum_{n=1}^{\infty}\frac{\kappa^n}{(n-1)!}[\partial_\tau {\mathcal G}{\scriptstyle [\tau]}, {\mathcal G}{\scriptstyle [\tau]}^{n-1},B]\no
&= \kappa [ Q_{\mathcal G\scriptscriptstyle [\tau]} V ,B]_{\mathcal G\scriptscriptstyle [\tau]}\no
&= -\kappa Q_{\mathcal G\scriptscriptstyle [\tau]} [ V ,B]_{\mathcal G \scriptscriptstyle [\tau]}+\kappa [ V ,Q_{\mathcal G\scriptscriptstyle [\tau]}B]_{\mathcal G\scriptscriptstyle [\tau]}.
\end{align}
Introducing a linear map $\widehat V{\scriptstyle[\tau]} : \mathcal{H} \rightarrow \mathcal{H}$ defined by 
\begin{align}
\widehat V{\scriptstyle[\tau]} := \kappa [ V , \hspace{3mm} ]_{\mathcal{G}\scriptscriptstyle[\tau] },
\end{align}
the above equation can be written as
\begin{align}
\partial_\tau Q_{\mathcal G\scriptscriptstyle [\tau]} = [\![ \widehat V{\scriptstyle [\tau]} ,Q_{\mathcal G\scriptscriptstyle [\tau]}]\!].
\label{Q G DE}
\end{align}

Since (\ref{Q G DE}) is of the form of (\ref{general DE}), 
the $\mathcal G$-shifted BRST operator $Q_{\mathcal G}$ 
can also be obtained by the similarity transformation from $Q$:
\begin{align}
Q_{\mathcal G}={\mathcal E_{V}} Q {\mathcal E_{V}}^\dagger.\label{QG def}
\end{align}
Here ${\mathcal E_{V}}$ is defined by path ordered exponential of $\widehat V$
\begin{align}
{\mathcal E_{V}}{\scriptstyle[\tau]}= \overset{\rightarrow}{\mathcal P} \exp \left( {\int^\tau_0 d\tau' \widehat V{\scriptstyle[\tau']} }\right),
\end{align}
and its explicit form is given by
\begin{align}
\mathcal E_V{\scriptstyle[\tau]}(A)=&\: A+\kappa\tau [V,A]+\frac{\kappa^2\tau^2} 2 \Big([V,QV,A]+[V,[V,A]]\Big)\no
&+\frac{\kappa^3\tau^3}6\Big( [V,QV,QV,A]+[V,[V,QV],A]+2[V,QV,[V,A]]\no
&\hspace{140pt}+[V,[V,QV,A]]+[V,[V,[V,A]]]\Big)+\cdots.
\end{align}
Since $\widehat V{\scriptstyle[\tau]}$ is BPZ odd,  ${\mathcal E_{V}}^\dagger$ is given by
\begin{align}
{\mathcal E_{V}} ^\dagger
=\overset{\leftarrow}{\mathcal P} \exp \left(-\int_0^1 d\tau \widehat V{\scriptstyle[\tau]}\right),
\end{align}
and it satisfies 
\begin{align}
{\mathcal E_{V}}^\dagger{\scriptstyle[\tau]}&={\mathcal E_{V}}{\scriptstyle[\tau]}^{-1}.
\end{align}
The nilpotency and cyclicity of $Q_{\mathcal G}$ follows from that of $Q$, which can be seen easily in the form using the similarity transformation:
\begin{align}
Q_{\mathcal G}^2 
=&{\mathcal E_{V}} Q {\mathcal E_{V}}^\dagger{\mathcal E_{V}} Q {\mathcal E_{V}}^\dagger
={\mathcal E_{V}} Q  Q {\mathcal E_{V}}^\dagger
=0,\no
(Q_{\mathcal G})^\dagger 
=&({\mathcal E_{V}} Q {\mathcal E_{V}}^\dagger)^\dagger =
{\mathcal E_{V}} Q^\dagger  {\mathcal E_{V}}^\dagger=
-Q_{\mathcal G}.
\end{align}

The equivalence of ${\mathcal E_{V}} Q {\mathcal E_{V}}^\dagger$ and $Q_{\mathcal G}$ which is defined as the $\mathcal G$-shifted operator
can be shown also by solving the differential equation for
$I{\scriptstyle[\tau]}={\mathcal E_{V}}^\dagger{\scriptstyle[\tau]} Q_{\mathcal G\scriptscriptstyle[\tau]} {\mathcal E_{V}}{\scriptstyle[\tau]}-Q$.
Differentiating $I{\scriptstyle[\tau]}$ by $\tau$, we obtain
\begin{align}
\partial_\tau I{\scriptstyle[\tau]}
=-{\mathcal E_{V}}^\dagger{\scriptstyle[\tau]} [V, Q_{\mathcal G\scriptscriptstyle[\tau]} {\mathcal E_{V}}{\scriptstyle[\tau]}\:\cdot\:]_\mathcal G
+{\mathcal E_{V}}^\dagger{\scriptstyle[\tau]} [Q_{\mathcal G\scriptscriptstyle[\tau]} V,{\mathcal E_{V}}{\scriptstyle[\tau]}\:\cdot\:]_\mathcal G
+{\mathcal E_{V}}^\dagger{\scriptstyle[\tau]} Q_{\mathcal G\scriptscriptstyle[\tau]} [V, {\mathcal E_{V}}{\scriptstyle[\tau]\:\cdot\:}]_\mathcal G
=0,
\end{align}
where we used $\partial_\tau Q_{\mathcal G\scriptscriptstyle[\tau]} =[ Q_{\mathcal G\scriptscriptstyle[\tau]}V, \:\cdot\:]_{\mathcal G\scriptscriptstyle[\tau]}$
and that $Q_\mathcal G$ act as a derivation on $[ \:\:, \:\:]_\mathcal G$, which follows from the $L_\infty$ relation of $\mathcal G$-shifted products.
Since the initial condition is $I{\scriptstyle[0]}=0$, we have $I{\scriptstyle[\tau]}=0$ for arbitrary $\tau$, and therefore 
${\mathcal E_{V}}^\dagger{\scriptstyle[\tau]} Q_{\mathcal G\scriptscriptstyle[\tau]} {\mathcal E_{V}}{\scriptstyle[\tau]}=Q.$
Acting ${\mathcal E_{V}}{\scriptstyle[\tau]}$ from the left and ${\mathcal E_{V}}^\dagger{\scriptstyle[\tau]}$ from the right,
we get (\ref{QG def}).

\section{Equivalence of on-shell conditions}

In this section, first, we explain that the similarity transformation of the BRST operator gives an invertible morphism of $L_{\infty }$-algebras preserving the zeros of the Maurer-Cartan element, which naturally induces a redefinition of string fields. 
Then, identifying these zeros with two on-shell conditions, we derive the explicit form of the field redefinition connecting two string fields, which guarantees the on-shell equivalence of two formulations. 

\subsection{$L_{\infty }$-morphism}

Let $( \mathcal{H} , {\bf L}  )$ and $( \mathcal{H^{\prime }}  , {\bf L}^{\prime } )$ be $L_{\infty }$-algebras, and ${\sf f}:S(\mathcal{H}) \rightarrow S(\mathcal{H}^{\prime })$ be a cohomomorphism of symmetric tensor algebras. 
A cohomomorphism ${\sf f} = \{ {\sf f}_{n} \} _{n=1}^{\infty }$ satisfying 
\begin{align}
\label{L morphism}
{\sf f } \,  {\bf L } = {\bf L}^{\prime } \, {\sf f}  
\end{align}
is called an {\it $L_{\infty }$-morphism}. 
For $L_{\infty }$-morphism ${\sf f} : \big( \mathcal{H} , {\bf L} \big) \rightarrow \big( \mathcal{H^{\prime }} , {\bf L}^{\prime } \big)$, one can get a relation in $\bigoplus _{k=1}^{n}\mathcal{H}^{\prime \wedge k}$ by evaluating the condition (\ref{L morphism}) with $\Phi _{1}\wedge \dots \wedge \Phi _{n} \in \mathcal{S(H)}$. 
For example,  
\begin{align}
&\sum_{i \leq n} \sum_{k_{1} < \dots < k_{i}} L_{i}^{\prime } \big( {\sf f}_{k_{1}} ( \Phi _{1} , \dots , \Phi _{k_{1}}) , {\sf f}_{k_{2}-k_{1}} ( \Phi _{k_{1} +1} , \dots , \Phi _{k_{2}} ) , \dots , {\sf f}_{k_{i} -k_{i-1}} ( \Phi _{k_{i-1}+1} , \dots , \Phi _{n} ) \big) 
\no 
& \hspace{10mm} = \sum_{k+l=n+1} \sum_{j=0}^{k-1} (-)^{\Phi _{1} + \dots + \Phi _{j}} {\sf f}_{k} \big(  \Phi _{1} , \dots , \Phi _{j} , L_{l} ( \Phi _{j+1} , \dots , \Phi _{j+l} ) , \Phi _{j+l+1} , \dots , \Phi _{n} \big) 
\end{align}
holds for $1\leq k_{1}, \dots ,k_{i}$ and $k_{i} = n$ in $\mathcal{H}^{\prime \wedge 1}$, whose first two relations are given by 
\begin{align}
L_{1}^{\prime } \big( {\sf f}_{1} ( \Phi _{1} ) \big) & = {\sf f}_{1} \big( L_{1} ( \Phi _{1} ) \big) 
\no  
L_{2}^{\prime } \big( {\sf f}_{1} ( \Phi _{1} ), {\sf f}_{1} (\Phi _{1} ) \big) + L_{1}^{\prime } \big( {\sf f}_{2} ( \Phi _{1} , \Phi _{2} ) \big) & =  {\sf f}_{2} \big( L_{1} (\Phi _{1}) , \Phi _{2} \big) + (-)^{\Phi _{1}} {\sf f}_{2} \big( \Phi _{1} , L_{1}( \Phi _{2} ) \big) 
\no \nonumber 
& \hspace{15mm} + {\sf f}_{1} \big( L_{2} ( \Phi _{1} , \Phi _{2} ) \big) .
\end{align}

Suppose that a set of multilinear maps ${\sf A} = \{ {\sf A}_{n} \} _{n=1}^{\infty }$ generates the similarity transformation of the BRST operator $Q$
\begin{align}
{\bf Q}^{\prime } :=  {\sf A} \, {\bf Q} \, {\sf A}^{\dagger } , 
\end{align}
where $Q :\mathcal{H} \rightarrow \mathcal{H}$, ${\sf A}_{n}: \mathcal{H}^{\wedge n} \rightarrow \mathcal{H}^{\prime }$, and $Q^{\prime } :\mathcal{H}^{\prime } \rightarrow \mathcal{H}^{\prime }$. 
By identifying ${\sf A} = \{ {\sf A}_{n} \} _{n=1}^{\infty }$ with a cohomomorphism on $\mathcal{S(H)}$ as shown in (\ref{cohom}), this ${\sf A}$ naturally gives the $L_{\infty }$-morphism ${\sf A} : ( \mathcal{H} , {\bf Q} ) \rightarrow (\mathcal{H}^{\prime }, {\bf Q}^{\prime } )$ because of 
\begin{align}
{\sf A} \, {\bf Q} = ( {\sf A} \, {\bf Q} \, {\sf A}^{\dagger } ) \, {\sf A} = {\bf Q}^{\prime } \, {\sf A}. 
\end{align}
Therefore, once the similarity transformation is given, one can consider the corresponding $L_{\infty }$-morphism, which becomes a powerful tool to discuss the on-shell equivalence. 

\niu{Field redefinition induced by $L_{\infty }$-morphism}

An $L_{\infty }$-morphism ${\sf f}: (\mathcal{H} , {\bf L} ) \rightarrow (\mathcal{H}^{\prime } , {\bf L}^{\prime })$ naturally induces a {\it formal field redefinition} from original string fields $\Phi \in \mathcal{H}$ to the new string field $\Phi ^{\prime }\in \mathcal{H}^{\prime }$ as follows 
\begin{align}
\Phi ^{\prime } := {\boldsymbol \pi } \big( {\sf f} (e^{\wedge \Phi }) \big) = \sum_{n=1}^{\infty } \frac{1}{n!}{\sf f}_{n} ( \overbrace{\Phi , \dots , \Phi }^{n} ) , 
\end{align}
which gives a nonlinear correspondence of two fields $\Phi \in \mathcal{H}$ and $\Phi ^{\prime } \in \mathcal{H}^{\prime }$.  
The word `formal' means that in general, these two state spaces of fields may not be equivalent: $\mathcal{H} \not= \mathcal{H}^{\prime }$. 
Then, by construction, it satisfies ${\sf f} \big( e^{\wedge \Phi } \big) = e^{\wedge \Phi ^{\prime }}$, which means that the formal field redefinition $\Phi ^{\prime } := {\boldsymbol \pi } \, {\sf f} (e^{\wedge \Phi }) $ %induced by an $L_{\infty }$-morphism ${\sf f}$ 
maps the group-like element $e^{\wedge \Phi }$ of the original $L_{\infty }$-algebra $(\mathcal{H} , {\bf L})$ to the group-like element $e^{\wedge \Phi ^{\prime }}$ of the new $L_{\infty }$-algebra $(\mathcal{H}^{\prime } , {\bf L}^{\prime })$. 
Furthermore, we quickly find that 
\begin{align}
\label{preserving MC zeros}
{\sf f} \, {\bf L} ( e^{\wedge \Phi } ) = {\bf L}^{\prime } \, {\sf f} (e^{\wedge \Phi } ) = {\bf L}^{\prime } ( e^{\wedge \Phi ^{\prime }}) , 
\end{align}
and therefore ${\sf f}$ maps the zeros of ${\bf L}(e^{\wedge \Phi } )$ to that of ${\bf L}^{\prime } ( e^{\wedge \Phi ^{\prime }} )$ unless $\Phi ^{\prime } =0$. 

\vspace{2mm}

Since utilizing the projector ${\boldsymbol \pi}: \mathcal{S(H)} \rightarrow \mathcal{H}$ defined in (\ref{pi}), one can write 
\begin{align} 
{\bf L} (e^{\wedge \Phi }) =  \big( {\boldsymbol \pi } \, {\bf L}(e^{\wedge \Phi } ) \big) \wedge e^{\wedge \Phi } , 
\end{align}
the Maurer-Cartan equation ${\boldsymbol \pi } \, {\bf L} ( e^{\wedge \Phi }) = 0$ is equivalent to ${\bf L} ( e^{\wedge \Phi }) = 0$. 
(Note that ${\bf L}(e^{\wedge \Phi } ) \in \mathcal{S(H)}$ and ${\boldsymbol \pi } \, {\bf L} ( e^{\wedge \Phi } ) \in \mathcal{H}$.) 
Hence, at least formally,  an $L_{\infty }$-morphism preserves the solutions of the Maurer-Cartan equations ${\boldsymbol \pi } \, {\bf L} ( e^{\wedge \Phi } ) = 0$ and ${\boldsymbol \pi } \, {\bf L}^{\prime } ( e^{\wedge \Phi ^{\prime }} ) = 0$, namely, on-shell states of string field theory.

\niu{Similarity transformation ${\bf G} \, {\bf Q} \, {\bf G}^{\dagger } = {\bf L}$}

The exponential map ${\bf G}$ defined in (\ref{G}) is a typical example of the above $L_{\infty }$-morphism. 
Let us consider the free theory in the small Hilbert space $\mathcal{H}_{\rm small}$ and let $(\mathcal{H}_{\rm small} , {\bf Q} )$ be the corresponding $L_{\infty }$-algebra. 
Recall that the Maurer-Cartan element ${\boldsymbol \pi } \, {\bf Q} \, (e^{\wedge \Phi })$ gives the equation of motion $Q \Phi = 0$, where $\Phi \in \mathcal{H}_{\rm small}$ is a closed NS string field belonging to the small Hilbert space $\mathcal{H}_{\rm small}$. 
Then, ${\bf G}$ gives a multilinear map ${\sf G}_{n} : \mathcal{H}_{\mathrm{small}}^{\wedge n} \rightarrow \mathcal{H}^{\prime }$ for $n \geq 1$. 

\vspace{2mm}

First, we can lift the set of (degree zero) multilinear maps ${\bf G}=\{ {\sf G}_{n} \} _{n=1}^{\infty }$ to the cohomomorphism of two Fock spaces $\mathcal{S(H}_{\rm small})$ and $\mathcal{S(H}^{\prime })$ as shown in (\ref{cohom}).
Then, by construction of the NS string products ${\bf L}$, the invertible cohomomorphism ${\bf G}$ satisfies 
\begin{align} 
{\bf G} \, {\bf Q} = {\bf L} \, {\bf G}, 
\end{align} 
which is the condition of the $L_{\infty }$-morphism ${\bf G}:(\mathcal{H}_{\rm small} , {\bf Q}) \rightarrow (\mathcal{H}^{\prime } , {\bf L})$. 
This ${\bf G}$, therefore, induces the following {formal} field redefinition 
\begin{align}
\Phi ^{\prime } := {\boldsymbol \pi } \big( {\bf G} (e^{\wedge \Phi }) \big) = \sum_{n=1}^{\infty } \frac{1}{n!}{\sf G}_{n}(\overbrace{\Phi , \dots , \Phi }^{n} ) .  
\end{align}
Note that $\eta \, \Phi ^{\prime } \not= 0$ and the state space $\mathcal{H}^{\prime }$ of $\Phi ^{\prime }$ is different from $\mathcal{H}_{\mathrm{small}}$. 
Hence, the resulting theory $\big( \mathcal{H^{\prime }} , {\bf L} \big)$, which is equivalent to the free theory $\big( \mathcal{H_{\mathrm{small}}} , {\bf Q} \big) $, is not equivalent to the $L_{\infty }$-type interaction theory $\big( \mathcal{H_{\mathrm{small}}} , {\bf L} \big) $ proposed by \cite{Erler:2014eba}. 
It clarifies the difference to compare the equation of motion in {\bf G}-mapped free theory 
\begin{align}
{\boldsymbol \pi } \, {\bf L} \big( e^{\wedge \Phi ^{\prime }} \big) = {\boldsymbol \pi } \, {\bf L} \big( {\bf G} (e^{\wedge \Phi }) \big) = {\boldsymbol \pi } \, {\bf G} \, {\bf Q} \, (e^{\wedge \Phi }) = 0
\end{align}
with that in the $L_{\infty }$-type interaction theory proposed by \cite{Erler:2014eba} 
\begin{align}
{\boldsymbol \pi } \, {\bf L} \big( e^{\wedge \Phi } \big) =  0 . 
\end{align}

\subsection{On-shell equivalence}

Combining the result of section 3 and the concept of $L_{\infty }$-morphism giving $(\ref{preserving MC zeros})$, we discuss the on-shell equivalence of $L_{\infty }$-type and WZW-like formulations. 

\niu{Similarity transformation ${\sf F} \, {\bf L} \, {\sf F}^{\dagger } = Q_{\mathcal{G} }$}

In section 3, we found that utilizing invertible maps ${\bf G}$ and $\mathcal{E}_{V}$, one can construct the NS string products ${\bf L}$ 
\begin{align}
{\bf L} &= {\bf G} \, {\bf Q} \, {\bf G}^{\dagger }  
\end{align}
and the $\mathcal{G}$-shifted BRST operator $Q_{\mathcal{G}}$ 
\begin{align}
Q_{\mathcal{G}} &= \mathcal{E}_{V} \, Q  \, {\mathcal{E}_{V}}^{\! \dagger } , 
\end{align}
which are similarity transformations of the BRST operator $Q$ respectively. 
Hence, composing ${\bf G}^{\dagger } = \{ {\sf G}^{\dagger }_{n} \} _{n=1}^{\infty }$ with $\mathcal{E}_{V}$, we can construct the invertible map ${\sf F} = \{ {\sf F}_{n}  \} _{n=1}^{\infty }$  
\begin{align}
{\sf F} := \mathcal{E}_{V} \, {\bf G}^{\dagger }  . 
\end{align} 
Note that we use $\mathcal{E}_{V}$ as a linear map and define the $n$-fold multilinear map ${\sf F}_{n} : = \mathcal{E}_{V} \, {\sf G}_{n}^{\dagger }$.  
This set of multilinear maps ${\sf F}= \{ {\sf F}_{n} \} _{n=1}^{\infty }$ is the key ingredient in deriving our main result. 
We can quickly find that ${\sf F}$ generates the similarity transformation connecting 
${\bf L}$ and 
$Q_{\mathcal{G}}$: 
\begin{align}
\nonumber 
{\sf F} \, {\bf L} \, {\sf F}^{\dagger } & = ( \mathcal{E}_{V} {\bf G}^{\dagger } ) \, {\bf G} \, {\bf Q} \, {\bf G}^{\dagger } \, ( {\bf G} \, {\mathcal{E}_{V}}^{\dagger } ) 
\no
&= \mathcal{E}_{V} \,  Q \, {\mathcal{E}_{V}}^{\! \dagger } 
\no 
&= Q_{\mathcal{G}} . 
\end{align} 

We write $\mathcal{H}^{\prime }$ for a state space on which $Q_{\mathcal{G}}$ acts as a linear nilpotent operator and consider two $L_{\infty }$-algebras $(\mathcal{H}^{\prime } , Q_{\mathcal{G}})$ and $(\mathcal{H}_{\rm small} , {\bf L} )$. 
Since ${\sf F}$ is invertible and ${\sf F} \, {\bf L} = Q_{\mathcal{G}} \, {\sf F}$ holds, this cohomomorphism ${\sf F}$, as well as ${\bf G}$, becomes an $L_{\infty }$-morphism ${\sf F} : (\mathcal{H}_{\rm small } , {\bf L} ) \rightarrow (\mathcal{H}^{\prime } , Q_{\mathcal{G}} )$. 
Then, we can consider the field redefinition induced by ${\sf F}$
\begin{align}
\label{Phi prime}
\Phi ^{\prime } := {\boldsymbol \pi } \, {\sf F} ( e^{\wedge \Phi } ) = \sum_{n=1}^{\infty } \frac{1}{n!}{\sf F}_{n} ( \overbrace{\Phi , \dots , \Phi }^{n})  
\end{align}
which provides the equivalence of two Maurer-Cartan equations ${\boldsymbol \pi } \, {\bf L} (e^{\wedge \Phi })=0$ and $Q_{\mathcal{G}} \Phi ^{\prime } = 0$ of two $L_{\infty }$-algebras $(\mathcal{H}_{\rm small} , {\bf L})$ and $(\mathcal{H}^{\prime } , Q_{\mathcal{G}} )$. 
The state space $\mathcal{H}^{\prime }$ is spanned by $\Phi ^{\prime }$. 

\niu{Equivalence of on-shell conditions}

We derive the on-shell equivalence of $L_{\infty }$-type and WZW-like formulations by identifying the on-shell states of WZW-like string field theory with the zeros of the Maurer-Cartan equation $Q_{\mathcal{G}} \, \Phi ^{\prime } = 0$, which provides the field redefinition of two string fields. 
While the equation of motion in $L_{\infty }$-type formulation is given by 
\begin{align}
\label{EKS eom}
\mathcal{F}_{\Phi } \equiv {\boldsymbol \pi } \, {\bf L} ( e^{\wedge \Phi })  = 0  
\end{align} 
using the group-like element $e^{\wedge \Phi }$ consisting of small-space NS string fields $\Phi \in \mathcal{H}_{\rm small}$, 
the equation of motion (or on-shell condition) in WZW-type formulation is given by 
\begin{align}
\label{WZW eom}
\mathcal{F}_{V} \equiv Q_{\mathcal{G}} \Psi _{\eta } = 0 , 
\end{align} 
where the field $\Psi _{\eta }$ is the associated field defined by $(\ref{associated field})$ with $\mathbb{X} = \eta $, a function of large-space NS string fields $V$. 
Since ${\bf L}^{2} = 0$, the state $\mathcal{F}_{\Phi }$ belongs to the kernel of ${\bf L}(e^{\wedge \Phi } \wedge \mathbb{I} )$, which is the generator of the gauge invariance of the $L_{\infty }$-type action. 
It is also clear that the state $\mathcal{F}_{V}$ is $Q_{\mathcal{G}}$-exact and belongs to the kernel of $Q_{\mathcal{G}}$, which generates a gauge invariance of the WZW-like action. 
In addition, since $\Psi _{\eta }$ satisfies $Q_{\mathcal{G}} \Psi _{\eta } = - \eta \, \mathcal{G}$, as well as the state $\mathcal{F}_{\Phi }$, the state $\mathcal{F}_{V}$ also belongs to the kernel of $\eta $, namely, the small Hilbert space $\mathcal{H}_{\rm small}$. 

\vspace{3mm}

Therefore, to obtain the equivalence of these on-shell conditions, we identify $\Phi ^{\prime }$ defined by $(\ref{Phi prime})$ with the associated field $\Psi _{\eta }$ 
\begin{align}
\label{identification}
\Phi ^{\prime } := {\boldsymbol \pi } \, {\sf F} ( e^{\wedge \Phi } ) \cong \Psi _{\eta }  
\end{align}
and $\mathcal{H}^{\prime }$ with the state space in which $\Psi _{\eta }$ lives.
It gives the field redefinition of two string fields and provides the on-shell equivalence of two formulations. 
Note, however, that there exist two ambiguities in the identification: the terms being equivalent to zero on the mass shell and the ${\bf L}$- or $Q_{\mathcal{G}}$-exact terms. 
In other words, $(\ref{identification})$ is the identification of two representatives $[\Phi ^{\prime }]$ and $[\Psi _{\eta }]$ of the equivalent class defined by the equivalence relation “belongs to the kernel of $Q_{\mathcal{G}}$ when ${\boldsymbol \pi } \, {\bf L} ( e^{\wedge \Phi } ) = 0$ holds”. 
The reason is that we used an $L_{\infty }$-morphism to derive $(\ref{identification})$. 
Since the $L_{\infty }$-morphism ${\sf F} : ( \mathcal{H}_{\mathrm{small}}  , {\bf L} ) \rightarrow ( \mathcal{H}^{\prime } , Q_{\mathcal{G}} )$ satisfies ${\sf F} \, {\bf L} = Q_{\mathcal{G}} \, {\sf F} $ and the linear map $Q_{\mathcal{G}}$ commutes with the projector ${\boldsymbol \pi }$, we can quickly find 
\begin{align}
\nonumber 
{\boldsymbol \pi } \, {\sf F} \, {\bf L} ( e^{\wedge \Phi }) & = {\boldsymbol \pi } \, Q_{\mathcal{G}} \, {\sf F} ( e^{\wedge \Phi } )  
\no
&\cong  Q_{\mathcal{G} } \, \Psi _{\eta } . 
\end{align} 
%%%%%
As we saw in $(\ref{preserving MC zeros})$, the $L_{\infty }$-morphism ${\sf F}$ maps the solution space of ${\boldsymbol \pi } \, {\bf L} ( e^{\wedge \Phi } ) =0$ into that of $Q_{\mathcal{G}} \Psi _{\eta }  \cong Q_{\mathcal{G}} \Phi ^{\prime } = 0$, rather than the equation of motion itself, and only provides the equivalence of the zeros, namely, on-shell states. 
Hence, it does not affect the on-shell equivalence to add the terms which vanishes on the mass shell or is ${\bf L}$- or $Q_{\mathcal{G}}$-exact. 

\vspace{3mm}

Solving the nonlinear correspondence (\ref{identification}) of two string fields $\Phi $ and $V$ with respect to the coupling constant $\kappa $, one can explicitly check the equivalence of two on-shell conditions (\ref{EKS eom}) and (\ref{WZW eom}). 
Furthermore, as we will see in section 4.3, the identification $(\ref{identification})$ also provides the off-shell equivalence of two formulations at least up to quartic order of fields. 

\niu{Correspondence of two string fields} 

Before solving $(\ref{identification})$, we rewrite it into a more computable form. 
As we explain, the correspondence of two string fields ${\boldsymbol \pi} \, {\mathsf F} ( e^{\wedge \Phi } ) \cong \Psi _{\eta }$, which provides the equivalence of ${\bf L} ( e^{\wedge \Phi }) = 0$ and $Q_{\mathcal{G} } \Psi _{\eta } = 0$, can be rewritten as 
\begin{align}
\label{condition}
{\boldsymbol \pi } \big( {\bf G}^{\dagger } ( e^{\wedge \Phi } ) \big) \cong \int_{0}^{1} d\tau \,  {\mathcal{E}_{V}}^{\! \dagger } \hspace{-0.3mm} {\scriptstyle [\tau]} \big( \eta V \big) . 
\end{align}
While the left hand side is a function of string fields $\Phi $ and $\xi \Phi $, the right hand side is a function of string fields $V$ and $\eta V$. 
Each side of (\ref{condition}) can be constructed by invertible transformations of small-space states $e^{\wedge \Phi }$ or $\eta V$. 
For this purpose, we show that the following relation holds 
\begin{align}
\label{aux}
{\mathcal{E}_{V}}^{\! \dagger } \Psi _{\mathbb{X}} = \int_{0}^{1}d\tau \, {\mathcal{E}_{V}}^{\! \dagger } \hspace{-0.3mm} {\scriptstyle [\tau]} \big( \mathbb{X} V \big) . 
\end{align}
Then, by construction of ${\sf F}$, $(\ref{condition})$ follows from $(\ref{identification})$ and $(\ref{aux})$. 
Recall that for any $A \in \mathcal{H}$, the defining equation of the map ${\mathcal{E}_{V}}^{\! \dagger }$ is given by 
\begin{align}
\frac{\partial }{\partial \tau } {\mathcal{E}_{V}}^{\! \dagger } \hspace{-0.3mm} {\scriptstyle [\tau]} \big( A \big) = - {\mathcal{E}_{V }}^{\! \dagger } \hspace{-0.3mm} {\scriptstyle [\tau]} \big( [ V ,  A  ] _{\mathcal{G}{\scriptscriptstyle [\tau ]} } \big) 
\end{align}
with the initial condition ${\mathcal{E}_{V}}^{\! \dagger } \hspace{-0.3mm} {\scriptstyle [\tau = 0 ]} = \mathbb{I}$. 
Thus, we find that 
\begin{align}
\nonumber 
\frac{\partial }{\partial \tau } \Big( {\mathcal{E}_{V}}^{\! \dagger } \hspace{-0.3mm} {\scriptstyle [\tau]} \, \Psi _{\mathbb{X}} {\scriptstyle [\tau ]} \Big) &= \Big( \frac{\partial }{\partial \tau } {\mathcal{E}_{V}}^{\! \dagger } \hspace{-0.3mm} {\scriptstyle [\tau]} \Big) \, \Psi _{\mathbb{X}} {\scriptstyle [\tau ] }+ {\mathcal{E}_{V}}^{\! \dagger } \hspace{-0.3mm} {\scriptstyle [\tau]} \, \Big( \frac{\partial }{\partial \tau } \, \Psi _{\mathbb{X}} {\scriptstyle [\tau ]} \Big) 
\no 
&= - {\mathcal{E}_{V}}^{\! \dagger } \hspace{-0.3mm} {\scriptstyle [\tau]} \Big( \big[ V , \Psi _{\mathbb{X}} {\scriptstyle [\tau ]} \big] _{\mathcal{G} {\scriptscriptstyle [\tau ]}} \Big) + {\mathcal{E}_{V}}^{\! \dagger } \hspace{-0.3mm} {\scriptstyle [\tau]} \Big( \mathbb{X} V + \kappa \big[ V , \Psi _{\mathbb{X}} {\scriptstyle [\tau ]} \big] _{\mathcal{G} {\scriptscriptstyle [\tau ]}} \Big) 
\no 
& = {\mathcal{E}_{V}}^{\! \dagger } \hspace{-0.3mm} {\scriptstyle [\tau]} \big( \mathbb{X} V \big)  . 
\end{align}
Since $\Psi _{\mathbb{X}}{\scriptstyle [\tau = 0]} =0$, we obtain $(\ref{aux})$. 
In addition to this fact, the relation $(\ref{aux})$ also implies that using the invertible map $\mathcal{E}_{V}$ and the state $\mathbb{X} V$, one can construct the associated field $\Psi _{\mathbb{X}}$ as 
\begin{align}
\label{another rep} 
\Psi _{\mathbb{X}} = \mathcal{E}_{V}  \int_{0}^{1} d\tau \, {\mathcal{E}_{V}}^{\! \dagger } \hspace{-0.3mm} {\scriptstyle [\tau]} \, (\mathbb{X} V) . 
\end{align}
Hence, we can write WZW-like actions using the BRST operator, exponential maps, and string fields. 
(See appendix A.3.) 

\subsection{On the off-shell equivalence}

In the rest, we show that the identification $\Phi ^{\prime } \cong \Psi _{\eta }$ also provides the off-shell equivalence at least up to quartic order of fields. 
Note that the on-shell equivalence is guaranteed by 
\begin{align}
{\boldsymbol \pi } \, {\bf G}^\dagger \big( e^{\wedge \Phi (t)} \big) \cong \int_{0}^{1} d\tau \,  {\mathcal{E}_{V(t) }}^{\! \! \dagger } \hspace{-0.3mm} {\scriptstyle [\tau]} \big( \eta V(t) \big) , 
\end{align} 
which is the identification of representatives. 
Therefore, choosing a BRST-exact term $B = \sum \kappa ^{n} B_{n+1}$ and an on-shell vanishing term $E = \sum \kappa ^{n} E_{n+1}$, we can write the relation of two string fields in the state space (not in the equivalent class) 
\begin{align}
\label{relation}
{\boldsymbol \pi } \, {\bf G}^\dagger \big( e^{\wedge \Phi (t)} \big) = \int_{0}^{1} d\tau \,  {\mathcal{E}_{V(t) }}^{\! \! \dagger } \hspace{-0.3mm} {\scriptstyle [\tau]} \big( \eta V(t) \big) + \sum_{n=0}^{\infty } \kappa ^{n} B_{n+1} + \sum_{n=0}^{\infty } \kappa ^{n} E_{n+1} . 
\end{align}
When we obtain the relation of two string fields $V$ and $\Phi $ in the state space by determining $B$-terms, $E$-terms, and the form of ${\bf G}^{\dagger }$, a partial gauge fixing condition giving $L_{\infty }$-relation in WZW-like formulation naturally appears. 
For simplicity, in the following computation, we take ${\bf G}^{\dagger }$ obtained by $(\ref{EKS gauge 2})$ and $(\ref{EKS gauge 3})$, which is the gauge product used in \cite{Erler:2014eba}. 

\niu{The string field $V$ in terms of $\Phi $}

It is in fact straightforward to determine $B$- and $E$-terms of $(\ref{relation})$ and obtain the relation in the state space. 
Let us expand $V$ in powers of $\kappa $
\begin{align}
V = \sum_{n=0}^{\infty } \kappa ^{n} V^{\Phi }_{n+1} ,
\end{align}
and let $V^{\Phi }_{n+1}$ be a homogeneous function of $(n+1)$ string fields $\Phi $. 
We would like to write the string field $V$ in terms of $\Phi $. 
For this purpose, we impose the $\xi $-exact condition $\xi V = 0$. 
Since ${\sf G}^{\dagger }_{1} ( \Phi )= \Phi $, the relation at $O(\kappa ^{0})$ is given by 
\begin{align}
\eta V ^{\Phi }_{1}  = \Phi - B_{1} - E_{1} . 
\end{align} 
Note that we have to take $B_{1}$ and $E_{1}$ such that $\eta (B_{1} + E_{1})=0$ because $\eta V$ and $\Phi $ belong to the kernel of $\eta $. 
The simplest choice would be $B_{1} = E_{1} = 0$. 
Under $\xi V = 0$, we obtain $V^{\Phi }_{1} = \xi \Phi $. 
Then, utilizing these $\eta V^{\Phi }_{1} = \Phi $ and $V^{\Phi }_{1} = \xi \Phi $, the relation at $O(\kappa )$ is given by 
\begin{align}
\eta V^{\Phi }_{2} & = \frac{1}{3!} \Big( \xi [ \Phi , \Phi ] - 2 [ \xi \Phi . \Phi ] \Big) + \frac{1}{2} [ V^{\Phi }_{1} , \eta V^{\Phi }_{1} ] - B_{2} - E_{2} 
\no 
& = \frac{1}{3!} \Big(  \xi [ \Phi ,\Phi ] + [ \xi \Phi , \Phi ] \Big) - B_{2} - E_{2} . 
\end{align}
We, therefore, set $B_{2} =  E_{2} =0$ and obtain $V^{\Phi }_{2} = \frac{1}{3!} \xi [ \xi \Phi , \Phi ]$ under $\xi V = 0$. 
Similarly, using these $V^{\Phi }_{1} = \xi \Phi $ and $V^{\Phi }_{2} = \frac{1}{3!} \xi [ \xi \Phi , \Phi ]$, we obtain the relation at $O(\kappa ^{2})$ as follows 
\begin{align}
\eta V^{\Phi }_{3} & = \frac{\xi }{4!} \Big( 
\frac{1}{4} X [ \Phi , \Phi , \Phi ] + \frac{3}{4} [ \xi X \Phi , \Phi , \Phi ] + \frac{5}{3} [ \Phi , \xi [ \Phi , \Phi ] ] - \frac{4}{3} [ \Phi , [ \xi \Phi , \Phi ] ] \Big)
\no 
& \hspace{5mm} + \frac{1}{4!} \Big(
- \frac{3}{4} [ \xi X \Phi , \Phi , \Phi ] - \frac{3}{4} X [ \xi \Phi , \Phi , \Phi ] - \frac{3}{2} [ \xi Q \Phi , \xi \Phi , \Phi ] + \frac{5}{2} [ Q \xi \Phi , \xi \Phi , \Phi ] 
\no 
& \hspace{10mm} + 
\frac{1}{3} [ \xi \Phi , \xi [ \Phi , \Phi ] ] + \frac{4}{3} [ \Phi , \xi [ \xi \Phi , \Phi ] ]  - \frac{2}{3} [ \xi \Phi , [ \xi \Phi , \Phi ] ] \Big) 
 - B_{3} - E_{3} . 
\end{align}
To obtain the representative which realizes that the right hand side belongs to the kernel of $\eta$, $B_{3}$ and $E_{3}$ have to satisfy $4! \cdot \eta (B_{3} + E_{3}) = Q ( [ \xi \Phi , \Phi , \Phi ] + \xi [\Phi , \Phi , \Phi ] ) + [ \xi Q \Phi , \xi \Phi , \Phi ] - \xi [ Q \Phi , \Phi , \Phi ] - 6[ \xi \Phi , Q \Phi , \Phi ]$. 
We take the following choice of $B_{3}$ and $E_{3}$ 
\begin{align}
B_{3} & = - \frac{1}{4!} Q \xi [ \xi \Phi , \Phi , \Phi ] ,
\\ 
E_{3} & = - \frac{3}{4!} \Big( \xi [ \xi \Phi , Q \Phi , \Phi ] + [ \xi \Phi , \xi Q \Phi , \Phi ] \Big) .
\end{align}
As a result, with the condition $\xi V = 0$, we obtain $V = V^{\Phi }_{1} + \kappa V^{\Phi }_{2} + \kappa ^{2} V^{\Phi }_{3} + O(\kappa ^{3})$ as follows 
\begin{align}
\label{PGF}
V &= \xi \Phi + \frac{\kappa }{3!} \xi [ \xi \Phi , \Phi ] + \frac{\kappa ^{2}}{4!} \Big( \frac{3}{2} \xi [ \xi \Phi , \xi Q \Phi , \Phi ] + \frac{5}{2} \xi [ \xi \Phi , Q \xi \Phi , \Phi ] -\frac{3}{4}\xi [ \xi Q \xi \Phi , \Phi , \Phi ] 
\no 
& \hspace{5mm} + \frac{1}{4} \xi Q \xi  [ \xi \Phi , \Phi , \Phi ] + \frac{4}{3} \xi [ \xi [ \xi \Phi , \Phi ] , \Phi ] - \frac{2}{3} \xi [ [\xi \Phi , \Phi ] , \xi \Phi ] + \frac{1}{3} \xi [ \xi [ \Phi , \Phi ] , \xi \Phi ]  \Big) + O(\kappa ^{3}) . 
\end{align}

Actually, it just gives a partial gauge fixing condition reducing WZW-like action to $L_{\infty }$-type action. 
We can directly check that this correspondence of two string fields provides the equivalence of WZW-like and $L_{\infty }$-type actions. 

\niu{Partial gauge fixing}

We start with expanding the WZW-like action in powers of $\kappa $: 
\begin{align}
\label{up to 5}
S_{\scriptscriptstyle{\rm WZW}} & = S_{2} [ V ] + \kappa S_{3} [ V ] + \kappa ^{2} S_{4} [ V ] + O(\kappa ^{3} )
\no 
& = \frac{1}{2} \langle \eta V , Q V \rangle + \frac{\kappa }{3!} \langle \eta V , [QV , V] \rangle + \frac{\kappa ^{2}}{4!} \langle \eta V , \big[ QV , QV , V \big] + \big[ [ QV , V ] , V \big] \rangle + O(\kappa ^{3}) .
\end{align}
First, we briefly carry out the computation up to $O(\kappa ^{2})$. 
We consider $S_{2}[V] + \kappa S_{3}[V]$ and substitute $V = V^{\Phi }_{1} + \kappa V^{\Phi }_{2} + O(\kappa ^{2}) $ into $S_{2} [ V ] = \frac{1}{2} \langle \eta V , Q V \rangle $ and $\kappa S_{3} [V] = \frac{\kappa }{3!} \langle \eta V , [ Q V , V ] \rangle $. 
Then, we find that $S_{2} [ V ] = S^{\prime }_{2} [\Phi ] + \kappa S^{\prime }_{3} [\Phi ] + \frac{\kappa }{3!} \langle \xi \Phi , Q [ \xi \Phi , \Phi ] \rangle $ and $\kappa S_{3} [ V ] = - \frac{\kappa }{3!} \langle \xi \Phi , Q [ \xi \Phi , \Phi ] \rangle $, where $S^{\prime }_{2} [ \Phi ] = \frac{1}{2} \langle \xi \Phi , Q \Phi \rangle $ and $S^{\prime }_{3}[ \Phi ] = \frac{1}{3!} \langle \xi \Phi , X [ \Phi , \Phi ] \rangle $. 
Note that these $S^{\prime }_{2} [\Phi ]$ and $S^{\prime }_{3}[\Phi ]$ are nothing but the kinetic term and the three point vertex of the $L_{\infty }$-type action $S_{\scriptscriptstyle{\rm EKS}}$ respectively. 
Therefore, a partial gauge fixing $V = \xi \Phi + \frac{\kappa }{3! } \xi [ \xi \Phi , \Phi ] + O (\kappa ^{2}) $ reduces WZW-like action $S_{2}[V] + \kappa S_{3}[V] $ to $L_{\infty }$-type action $S^{\prime }_{2}[\Phi ] + \kappa S^{\prime }_{3}[\Phi ]$: 
\begin{align}
S_{\scriptscriptstyle{\rm WZW}} = S_{2}[V] + \kappa S_{3} [V] + O(\kappa ^{2}) 
\overset{(\ref{PGF})}{=} S^{\prime }_{2} [\Phi ] + \kappa S^{\prime }_{3} [\Phi ] + O(\kappa ^{2}) = S_{\scriptscriptstyle{\rm EKS}}.
\end{align}
In the same way, substituting $(\ref{PGF})$ into the WZW-like action $(\ref{up to 5})$, we obtain 
\begin{align}
S_{\scriptscriptstyle{\rm WZW} } &= \frac{1}{2} \langle \xi \Phi , Q \Phi \rangle + \frac{\kappa }{3!} \langle \xi \Phi , [ X \Phi , \Phi ] \rangle 
+ \frac{\kappa ^{2}}{4!} \langle  \xi \Phi , \frac{1}{3} [ \Phi , \xi X [ \Phi , \Phi ] \rangle  
\no 
& \hspace{4mm} + \frac{\kappa ^{2} }{4!} \langle \xi \Phi ,  \Big( \frac{5}{3} X [ \Phi , \xi [ \Phi , \Phi ] ] - \frac{3}{4} X [ \xi \Phi , [\Phi , \Phi ] ] +\frac{1}{6} X[ \Phi , [ \xi \Phi , \Phi ] ] \Big) \rangle 
\no & \hspace{10mm} 
+  \frac{\kappa ^{2}}{4!} \langle \xi \Phi , \Big( \frac{1}{4} [ X^{2} \Phi , \Phi , \Phi ] + \frac{3}{4} [ X\Phi , X \Phi , \Phi ]  \Big) \rangle 
+ O(\kappa ^{3}) . 
\end{align}
This is just equal to the $L_{\infty }$-type action: 
\begin{align}
S_{\scriptscriptstyle{\rm EKS} } &= S'_{2} [\Phi ] + \kappa S'_{3} [\Phi ] + \kappa ^{2} S'_{4} [\Phi ] + O(\kappa ^{3}) 
\no
&= \frac{1}{2} \langle \xi \Phi , Q \Phi \rangle + \frac{\kappa }{3!} \langle \xi \Phi , [ X \Phi , \Phi ] \rangle 
+ \frac{\kappa ^{2}}{4!} \langle  \xi \Phi , \frac{1}{3} [ \Phi , \xi X [ \Phi , \Phi ] \rangle  
\no 
& \hspace{4mm} + \frac{\kappa ^{2} }{4!} \langle \xi \Phi ,  \Big( \frac{5}{3} X [ \Phi , \xi [ \Phi , \Phi ] ] - \frac{3}{4} X [ \xi \Phi , [\Phi , \Phi ] ] +\frac{1}{6} X[ \Phi , [ \xi \Phi , \Phi ] ] \Big) \rangle 
\no & \hspace{10mm} 
+  \frac{\kappa ^{2}}{4!} \langle \xi \Phi , \Big( \frac{1}{4} [ X^{2} \Phi , \Phi , \Phi ] + \frac{3}{4} [ X\Phi , X \Phi , \Phi ]  \Big) \rangle + O(\kappa ^{3}) . 
\end{align}

Hence, the correspondence of fields $(\ref{PGF})$ derived from the identification $\Phi ^{\prime } \cong \Psi _{\eta }$ guaranteeing the on-shell equivalence matches two actions and provides the off-shell equivalence of two formulations at least up to quartic order. 
In the above, we use ${\bf G}^{\dagger }$ obtained by $(\ref{EKS gauge 2})$ and $(\ref{EKS gauge 3})$. 
Note, however, that there is another choice of ${\bf G}^{\dagger }$ since the solution of the differential equation providing the NS string products ${\bf L} = {\bf G} \, {\bf Q} \, {\bf G}^{\dagger }$ is not unique, and another choice of ${\bf G}^{\dagger }$ gives another NS string product satisfying $L_{\infty }$-relations and $\eta $-derivation properties \cite{Ph.D hm}. 
In that sense, it would be expected that by choosing partial gauge fixing conditions of the WZW-like action, we can obtain any corresponding $L_{\infty }$-type actions.

\section{Conclusion and discussion}

%[conclusion]
In this paper, we derive the condition providing the on-shell equivalence of $L_\infty$-type and WZW-like formulations.
We showed nonlinear nilpotent operators ${\bf L}$ and $Q_{\mathcal G}$ in both formulations
can be represented by the similarity transformations $\mathbf G$ and $\mathcal E_V$ of the BRST operator $Q$.
Utilizing them, we construct the similarity transformation $\mathsf F = \mathcal E_V \mathbf G^\dagger$ connecting ${\bf L}$ with $Q_{\mathcal{G}}$.
This $\mathsf F$ gives an $L_\infty$-morphism and 
naturally induces the field redefinition $\Phi' = \mathsf F(e^{\wedge\Phi})$ which preserves the zeros of ${\bf L} (e^{\wedge \Phi } )$ 
and $Q_{\mathcal{G}} \, \Phi ^{\prime }$. 
By the Identification $\Phi' \cong \Psi_\eta$
of these zeros with on-shell states,
one can obtain the equivalence of two on-shell conditions $\mathbf L (e^{\wedge\Phi})=0$ and $Q_\mathcal G \Psi_\eta=0$.

%\begin{figure}[ht]
\begin{center}
\begin{tikzpicture}
\node at (-4,0) {$L_\infty$};
\node at (-4,-2.4) {WZW};

\node at (-2,1.1) {Free theories};
\node[above] at (-2,0) {$\displaystyle Q\Phi =0 $};
\node[below] at (-2,0) {$(\Phi, Q)$};
\node[above] at (-2,-2.4) {$\displaystyle Q\eta V =0 $};
\node[below] at (-2,-2.4) {$(\eta V,Q)$};

\begin{scope}[xshift=0.3cm]
\draw[thick, to-to] (-1,0) -- (1,0);
\node[above] at (0,0.1) {replacing};
\draw[thick, to-to] (-1,-2.4) -- (1,-2.4);
\node[above] at (0,-2.3) {replacing};
\end{scope}

\node at (4.8,1.1) {Interacting theories};

\begin{scope}[xshift=4cm]
\node[above] at (3,0) {$\displaystyle \boldsymbol\pi\mathbf L (e^{\wedge\Phi})  =0$};
\node[below] at (3,0) {$\displaystyle (\Phi, \mathbf L)$};
\node[above] at (3,-2.4) {$\displaystyle Q_{\mathcal G} \Psi_\eta =0 $};
\node[below] at (3,-2.4) {$\displaystyle  (\Psi_\eta,Q_{\mathcal G}) $};
\end{scope}

\begin{scope}[xshift=5.1cm]
\draw[thick, -to] (-0.5,0.1) -- (0.5,0.1);
\draw[thick, to-] (-0.5,-0.1) -- (0.5,-0.1);
\node[above] at (0,0.1) {$\mathbf G$};
\node[below] at (0,-0.1) {$\mathbf G^\dagger $};
\draw[thick, -to] (-0.5,-2.3) -- (0.5,-2.3);
\draw[thick, to-] (-0.5,-2.5) -- (0.5,-2.5);
\node[above] at (0,-2.3) {$\mathcal E _V$};
\node[below] at (0,-2.5) {$\mathcal E _V^\dagger$};
\end{scope}

\begin{scope}[xshift=4.8cm]
\draw[thick, to-] (-0.1,-0.85) -- (-0.1,-1.6);
\draw[thick, -to] (0.1,-0.85) -- (0.1,-1.6);
\node [left] at (-0.1,-1.225) {$\mathsf F^\dagger$};
\node [right] at (0.1,-1.225) {$\mathsf F$};
\end{scope}

\begin{scope}[xshift=0cm]
\node[above] at (3,0) {$\displaystyle Q \boldsymbol\pi \mathbf G^\dagger (e^{\wedge\Phi}) =0$};
\node[below] at (3,0.1) {$\displaystyle (\mathsf G^\dagger (e^{\wedge\Phi}), Q)$};
\draw[rounded corners, dashed](1.5,-0.8) rectangle +(6.7,1.6);
\node[above] at (3,-2.4) {$  Q  \int\mathcal E_V^\dagger(\eta V)=0 $};
\node[below] at (3,-2.3) {$  ( \int\mathcal E_V^\dagger (\eta V), Q) $};
\draw[rounded corners, dashed](1.5,-3.25) rectangle +(6.7,1.6);
\end{scope}
\end{tikzpicture}
\end{center}
%\caption{}
%\end{figure}

In addition, we have confirmed up to quartic order
that this field redefinition $\Phi' \cong \Psi_\eta$
also provides the equivalence of two actions $S_{\scriptscriptstyle{\rm EKS}}$ and $S_{\scriptscriptstyle{\rm WZW}}$.
In particular,
the partial-gauge-fixing conditions giving $L_\infty$-relations in WZW-like string field theory can be obtained.
Although we showed them in the language of NS closed superstrings, our derivation works also for NS open superstrings and NS-NS closed superstrings.

%[Toward the off-shell equivalence]
Toward the equivalence of two formulations,
it is necessary to understand whether and how the on-shell equivalence condition $\Phi' \cong \Psi_\eta$ works off shell in all order.
As we explained in section 4, the on-shell equivalence condition $\Phi' \cong \Psi_\eta$,
or equivalently
\begin{align} 
{\boldsymbol \pi } \, {\mathbf G}^\dagger \big( e^{\wedge \Phi (t)} \big) & \cong \int _{0}^{1} d\tau \, {\mathcal{E}_{V(t)}}^{\! \! \dagger } \hspace{-0.3mm} {\scriptstyle [\tau]} \big( \eta V(t) \big) ,
\label{on equiv 1}
\end{align}
determines the correspondence of two string fields only up to $Q$-exact terms and terms which is equivalent to zero on shell.
We need to understand the off-shell prescription to solve (\ref{on equiv 1}) which leads to the equivalence of the actions in two formulations in all order,
namely the all-order partial-gauge-fixing condition giving $L_\infty$-relations in WZW-like string field theory.

\vspace{1mm}

%[Introduce App]
It is also suggestive that the interacting actions in both formulations can be written in the same form.
One can write the action in the $L_\infty$-type formulation as
\begin{align}
S_{\scriptscriptstyle{\rm EKS}} 
& = \int_{0}^{1} dt \, \langle {\boldsymbol \pi }  \boldsymbol{\xi }_{t}^{{\bf G}^\dagger}   {\bf G}^\dagger ( e^{\wedge \Phi (t)} ) ,  \,  Q \, {\boldsymbol \pi } \big({\bf G}^\dagger ( e^{\wedge \Phi (t)} ) \big) \rangle,
\end{align}
and the action in WZW-like formulation, under $\xi V=0$, as
\begin{align}
S_{\scriptscriptstyle \rm WZW}
&=\int^1_0 dt \langle  \xi_t^{({\scriptstyle \int} \mathcal E^\dagger)}  ({\textstyle \int} \mathcal E^\dagger) \eta V , Q ({\textstyle \int} \mathcal E^\dagger) \eta V\rangle,
\end{align}
where we denote $\int_0^1 d\tau \mathcal E_V^\dagger{\scriptstyle[\tau]}$ as $({\textstyle \int} \mathcal E^\dagger)$ for the notational simplicity,
and we defined $\mathsf f$-deformed $\xi_t$ by $\xi_t^{\mathsf f} := \mathsf f \, \xi_t \,\mathsf f^{-1}=\mathsf f \,  \partial_t \xi \,\mathsf f^{-1}$.
The property of $\mathbf G$ is well known in terms of the cyclic $L_\infty$-algebra, although the property of $\int \mathcal E$ is not yet well understood. 
For more detail, see appendix A.

\vspace{1mm}

%[Comments on Z_{2}-reversing] 
There is an alternative approach. 
When we can truncate the higher vertices $\{ L_{n}^{B} \} _{n=3}^{\infty }$ of bosonic string field theory,\footnote{Then, we can obtain a single covering of the bosonic moduli of Riemann surfaces by Feynman graph built using only the cubic vertex. } the WZW-like action reduces to the WZW-type one and the $\mathbb{Z}_{2}$-reversing symmetry enhances. 
The $\mathbb{Z}_{2}$-reversing changes the algebraic relations appearing in the WZW-like formulation by switching the role of some fields or operators, however, we can find that some algebraic relations appearing in $A_{\infty }/L_{\infty }$-type theory are equivalent to those in $\mathbb{Z}_{2}$-reversed WZW-type theory. 
This fact guarantees the off-shell equivalence of two theories\cite{Erler:2015rra, Erler:2015uba}. 
It would be interesting to seek a nonlinear extended version of this $\mathbb{Z}_{2}$-reversing operation.  

\vspace{1mm}

%[BV]
The construction of the all-order Batalin-Vilkovisky master action\cite{Batalin:1981jr} is one of the important problem\cite{Torii:2011,Kroyter:2012ni,Berkovits:2012np}.
Owing to the cyclic $L_\infty$-algebra, the classical BV master action in the small-space formulation is given by relaxing the ghost number of the string field.
On the other hand, that in the large-space formulation has not been constructed.
In addition to the equivalence of the actions, 
the all-order correspondence of the gauge transformations of string fields in two formulations will be also impotent.
To construct the quantum BV master action, the inclusion of the R sector will be necessary.
Although the complete action including the R sector has not been understood,
as the important step toward it, the equations of motions and pseudo actions are well discussed in \cite{Berkovits:2001im,Michishita:2004by,Kunitomo:2013mqa,Kunitomo:2014qla,Erler:talk}.
It would be interesting to understand them in terms of similarity transformations.

%\clearpage

\section*{Acknowledgement}

We would like to express our gratitude to the members of Komaba particle theory group, in particular, our supervisors, Mitsuhiro Kato and Yuji Okawa. 
We also would like to thank Shota Komatsu for comment on the draft of this paper. 
The work of H.M. was supported in part by Research Fellowships of the Japan Society for the Promotion of Science for Young Scientists. 

\vspace{1mm}

Finally, we would like to express our gratitude to the organizers of the conference ``String Field Theory 2015 Nara'' held in Japan and the organizers of the international conference on ``String Field Theory and Related Aspects VII'' held in China. 

\appendix 

\section{Cyclic $L_{\infty }$-algebras and actions} 

Toward the all-order off-shell equivalence of two formulations, we summarize some properties of actions for NS closed string field theory. 
It would be important to understand the compatibility of the (naive) cyclic property and the nilpotent property. 

\subsection{Cyclic $L_{\infty }$-morphism}

Let $( \mathcal{H} , {\bf L} , \omega )$ and $( \mathcal{H^{\prime } } , {\bf L}^{\prime } , \omega ^{\prime } )$ be cyclic $L_{\infty }$-algebras, and ${\sf f}: (\mathcal{H}, {\bf L})  \rightarrow (\mathcal{ H^{\prime }} , {\bf L}^{\prime } )$ be an $L_{\infty }$-morphism of $L_{\infty }$-algebras. 
An $L_{\infty }$-morphism ${\sf f} = \{ {\sf f}_{n} \} _{n=1}^{\infty }$ satisfying 
\begin{align} 
\langle \omega ^{\prime } | \, {\sf f}_{1} ( \Phi _{1} ) \otimes {\sf f}_{1} ( \Phi _{2} )  = \langle \omega | \, \Phi _{1} \otimes \Phi _{2} , 
\end{align}
and for fixed $n \geq 3$, 
\begin{align}
\sum_{\substack{k+l = n\\ k,l \geq 1}} {\sum_{\sigma }}' (-)^{|\sigma |} \langle \omega ^{\prime } | \, {\sf f}_{k} ( \Phi _{\sigma (1)} , \dots , \Phi _{\sigma (k)} ) \otimes {\sf f}_{l} (\Phi _{\sigma (k+1)} , \dots , \Phi _{\sigma (n)} ) = 0 
\end{align} 
is called a {\it cyclic} $L_{\infty }$-{\it morphism}, where ${}^{\forall }\Phi _{1}, \dots , \Phi _{n} \in \mathcal{H}$ and ${\sf f}_{k}:\mathcal{H}^{\wedge k} \rightarrow \mathcal{H}^{\prime }$. 

\vspace{2mm}

In closed (super-) string field theory, a cyclic $L_{\infty }$-morphism\footnote{Note that $f_{1}: \mathcal{H} \rightarrow \mathcal{H}^{\prime }$ may not be an isomorphism. } preserves the value of the action. 
The equality follows from the fact that the symplectic structures on both sides are nondegenerate.

\niu{Similarity transformation generated by {\bf G}}

Note that the $L_{\infty }$-morphism ${\bf G} = \{  {\sf G}_{n} \} _{n=1}^{\infty }$ is invertible ${\bf G} \, {\bf G}^{\dagger } = {\boldsymbol 1}$ . 
Acting ${\bf G} \, {\bf G}^{\dagger } = {\boldsymbol 1}$ on $\sum_{n} \Phi _{1} \wedge \dots \wedge \Phi _{n} \in \mathcal{S(H)}= \bigoplus _{n} \mathcal{H}^{\wedge n}$, we obtain 
\begin{align}
\sum_{n=k+l} {\sum_{\sigma }}'  {\sf G}_{k+1} \big( \Phi _{\sigma (1) } , \dots , \Phi _{\sigma (k) }, {{\sf G}^{\dagger }}_{l} ( \Phi _{\sigma (k+1)} , \dots , \Phi _{\sigma (n) }) \big) = \Phi _{1}  
\end{align} 
in $\mathcal{H}^{\prime \wedge 1}$. 
Let $\langle \Phi _{0} , \Phi _{1} \rangle $ be the BPZ inner product in the large Hilbert space $\mathcal{H}$. 
The above relation implies 
\begin{align} 
\langle \Phi _{0} , \Phi _{1} \rangle  = \sum_{n=k+l} {\sum_{\sigma }}' (-)^{|\sigma |} \langle {{\sf G}^{\dagger }}_{k+1} ( \Phi _{\sigma (0)} , \dots , \Phi _{\sigma (k)} ) , {{\sf G}^{\dagger }}_{n-k} ( \Phi _{\sigma (k+1)} , \dots , \Phi _{\sigma (n)} ) \rangle . 
\end{align}
Since ${\sf G}_{1} = {{\sf G}_{1} }^{\dagger } = {\boldsymbol 1}$ and the BPZ inner product $\langle \Phi _{0} , \Phi _{1} \rangle $ is nondegenerate, we obtain 
\begin{align}
\langle {\sf G}_{1} ( \Phi _{0} ) , {\sf G_{1}} ( \Phi _{1} ) \rangle = \langle \Phi _{0} , \Phi _{1} \rangle , 
\end{align}
and for fixed $n \geq 3$, 
\begin{align}
\sum_{\substack{k+l = n\\ k,l \geq 1}} {\sum_{\sigma }}' (-)^{|\sigma |} \langle  {{\sf G}^{\dagger }}_{k} ( \Phi _{\sigma (1)} , \dots , \Phi _{\sigma (k)} ) , \, {{\sf G}^{\dagger }}_{l} (\Phi _{\sigma (k+1)} , \dots , \Phi _{\sigma (n)} ) = 0 . 
\end{align}
Hence, ${\bf G}$ and ${\bf G}^{\dagger }$ become cyclic $L_{\infty }$-morphisms. 

A cyclic $L_{\infty }$-morphism preserves the value of actions for string field theory. 
Let us consider the free action $S_{2}$ using small-space closed NS string fields $\Phi \in \mathcal{H}_{\mathrm{small}}$. 
Introducing a real parameter $t \in [0,1]$, we can rewrite the action as follows  
\begin{align}
S_{2} & = \frac{1}{2} \langle \xi \Phi , Q \Phi \rangle 
\no 
& = \int_{0}^{1} dt \, \langle \xi \partial _{t} ( t\Phi ) , Q ( t \Phi ) \rangle 
\end{align}
Furthermore, in coalgebraic representation, $S_{2}$ becomes 
\begin{align}
S_{2} = \int_{0}^{1} dt \, \langle {\boldsymbol \pi } \, {\boldsymbol \xi}_{t} \big( e^{\wedge t\Phi } \big) , \,  {\boldsymbol \pi } \, {\bf Q}\big( e^{\wedge t \Phi } \big) \rangle , 
\end{align}
where ${\boldsymbol \pi } : \mathcal{S( H_{ \mathrm{small} } ) }  \rightarrow \mathcal{H}_{\mathrm{small} }$ is the one-state projector defined in (\ref{pi}) and ${\boldsymbol \xi}_{t}$ is the coderivation constructed from the linear operator $\xi \partial _{t} : \mathcal{H}_{\mathrm{small}} \rightarrow \mathcal{H}$ defined in (\ref{Xi t}). 
The cyclic $L_{\infty }$-morphism ${\bf G}^{\dagger } \, {\bf L} = {\bf Q} \, {\bf G}^{\dagger } $ preserves the action: 
\begin{align}
S_{2} & = \int_{0}^{1} dt \, \langle {\boldsymbol \pi } \, {\boldsymbol \xi}_{t} \big(  e^{\wedge t\Phi } \big) , \,  {\boldsymbol \pi } \, {\bf Q} \big( {{\bf G}}^{\dagger } \, {\bf G} ( e^{\wedge t \Phi } ) \big) \rangle 
\no 
& = \int_{0}^{1}dt \, \langle {\boldsymbol \pi } \, {\bf G} \big( {\boldsymbol \xi}_{t} ( e^{\wedge t\Phi } ) \big) , \,   {\boldsymbol \pi } \, {\bf L} \big( {\bf G} ( e^{\wedge t \Phi } ) \big) \rangle 
\end{align}
Using the ${\bf G}$-transformed coderivation ${\boldsymbol \xi}^{\sf G}_{t} := {\bf G} \, {\boldsymbol \xi }_{t} \, {\bf G}^{\dagger }$ and the ${\bf G}$-transformed field 
\begin{align}
{\sf G} ( t\Phi ) : = {\boldsymbol \pi } \, {\bf G} ( e^{\wedge t\Phi } ) = \sum_{n=1}^{\infty } \frac{1}{n!}{\sf G}_{n} ( \overbrace{\Phi , \dots , \Phi }^{n} ) , 
\end{align}
the free action is given by 
\begin{align}
S_{2} & = \int_{0}^{1} dt \, \langle  {\boldsymbol \pi } \, {\boldsymbol \xi}^{\sf G}_{t} \big( {\bf G} ( e^{\wedge t\Phi } ) \big) , \,   {\boldsymbol \pi } \, {\bf L} \big( {\bf G} ( e^{\wedge t \Phi } ) \big) \rangle 
\no 
& =  \int_{0}^{1} dt \, \langle  {\boldsymbol \pi } \, {\boldsymbol \xi}^{\sf G}_{t} \big( e^{\wedge {\sf G}( t\Phi )} \big) , \,   {\boldsymbol \pi } \, {\bf L} \big( e^{\wedge {\sf G} ( t \Phi ) } \big) \rangle 
\end{align}
Note that a cohomomorphism preserves the group-like elements ${\bf G} ( e^{\wedge \Phi } ) = e^{\wedge {\sf G} (\Phi ) }$. 
The free action $S_{2}$ consists of the triplet $( \Phi , {\bf Q} , {\boldsymbol \xi _{t}} )$, which naturally gives a cyclic $L_{\infty }$-algebra, 
and thus, the triplet $( \Phi , {\bf Q} , {\boldsymbol \xi }_{t})$ is mapped to $({\sf G}(\Phi ) , {\bf L} , {\boldsymbol \xi }^{\sf G}_{t} )$ by the cyclic $L_{\infty }$-morphism ${\bf G}$. 
This is just a field redefinition from $\Phi $ to ${\sf G} (\Phi )$, which transforms from the ${\boldsymbol A}$ to ${\bf G} \, {\boldsymbol A} \, {\bf G}^{\dagger }$. 

\subsection{$L_\infty$-type action}

In the previous subsection, we explain that the action can be characterized by the triplet giving a cyclic $L_{\infty }$-algebra.
In this subsection, we understand the $L_\infty$-type action using the triplet.

As explained in section 3, $L_{\infty }$-type action can be written as
\begin{align}
S_{\scriptscriptstyle{\rm EKS}} & = \int_{0}^{1} dt \, \langle {\boldsymbol \pi } ( {\boldsymbol \xi }_{t} \, e^{\wedge \Phi (t)} ) , {\boldsymbol \pi } \big( {\bf L}( e^{\wedge \Phi (t)} ) \big) \rangle . 
\end{align}
One can find that the action consists of the triplet $( \Phi , {\bf L} , {\boldsymbol \xi }_{t} )$.

Since ${\bf G}$ is the cyclic $L_{\infty }$-morphism satisfying ${\bf L} = {\bf G} \, {\bf Q} \, {\bf G}^{\dagger }$,
we can transform the action into the following form:
\begin{align} 
S_{\scriptscriptstyle{\rm EKS}} & = \int_{0}^{1} dt \, \langle {\boldsymbol \pi } ( \boldsymbol{\xi }_{t} \, e^{\wedge \Phi (t)} ) , \, {\boldsymbol \pi } \big( {\bf G} \, {\bf Q} \, {\bf G}^\dagger  ( e^{\wedge \Phi (t)} ) \big) \rangle 
\no 
& = \int_{0}^{1} dt \, \langle {\boldsymbol \pi } \big( {\bf G}^\dagger ( \boldsymbol{\xi }_{t} e^{\wedge \Phi (t)} ) \big)  ,  \, {\boldsymbol \pi }\,  \mathbf Q \big({\bf G}^\dagger ( e^{\wedge \Phi (t)} ) \big) \rangle 
\no 
& = \int_{0}^{1} dt \, \langle {\boldsymbol \pi }  \boldsymbol{\xi }_{t}^{{\bf G}^\dagger}   {\bf G}^\dagger ( e^{\wedge \Phi (t)} ) , 
 \,  {\boldsymbol \pi } \mathbf Q \, \big({\bf G}^\dagger ( e^{\wedge \Phi (t)} ) \big) \rangle 
\no
& = \int_{0}^{1} dt \, \langle {\boldsymbol \pi }  \boldsymbol{\xi }_{t}^{{\sf G}^\dagger} ( e^{\wedge {\sf G}^\dagger(\Phi (t))} ) , 
 \,  {\boldsymbol \pi } \mathbf Q \, (  e^{\wedge {\sf G}^\dagger (\Phi (t) )}  ) \rangle . \label{action_eks_form1}
\end{align}
Thus, one can find that the action can be represented
also by $({\sf G}^{\dagger } (\Phi ),   \mathbf Q, {\boldsymbol \xi }^{{\sf G}^{\dagger }}_{t})$.
In other words, 
the triplet $( \Phi , {\bf L} , {\boldsymbol \xi }_{t} )$ is mapped to 
$({\sf G}^{\dagger } (\Phi ),   \mathbf Q, {\boldsymbol \xi }^{{\sf G}^{\dagger }}_{t})$
by the cyclic $L_\infty$-morphism $\mathsf G$,
which preserves the value of the action since $\mathbf G \mathbf G^\dagger = \1$.

It is interesting to compare this with the free action $S_2$ which consists of $( \Phi , {\bf Q} , {\boldsymbol \xi _{t}} )$.
One can easily find that the interacting theory can be obtained from the free theory
by {\it replacing} the field $\Phi $ and ${\boldsymbol \xi }_{t}$ by ${\sf G}^{\dagger } (\Phi )$ and ${\boldsymbol \xi }^{{\sf G}^{\dagger }}_{t}$ at once.
It can be understood by the following table.

%\begin{figure}[h]
\begin{center}
\begin{tikzpicture}[xscale =0.95 , yscale =0.9]

\begin{scope}[xshift=10]
\node[above] at (-5,1.5) {Free theory};

\node [above]at (-5,0) {$\displaystyle
\int_{0}^{1} dt \, \langle {\boldsymbol \pi } \, {\boldsymbol \xi}_{t} \big( e^{\wedge t\Phi } \big) , \,  {\boldsymbol \pi } \, {\bf Q}\big( e^{\wedge t \Phi } \big) \rangle$};
\node [below] at (-5,0) {$\displaystyle( \Phi , {\bf Q} , {\boldsymbol \xi _{t}} )$};

\draw[thick, -to] (-5.1,-0.8) -- (-5.1,-1.8);
\draw[thick, to-] (-4.9,-0.8) -- (-4.9,-1.8);
\node [left] at (-5.1,-1.2) {$\mathbf G$};
\node [right] at (-4.9,-1.2) {$\mathbf G^\dagger$};

\node [above]at (-5,-3) {$\displaystyle
\int_{0}^{1} dt \, \langle  {\boldsymbol \pi } \, {\boldsymbol \xi}^{\sf G}_{t} \big( e^{\wedge {\sf G}( t\Phi )} \big) , \,   {\boldsymbol \pi } \, {\bf L} \big( e^{\wedge {\sf G} ( t \Phi ) } \big) \rangle $};
\node [below] at (-5,-3) {$\displaystyle({\sf G}(\Phi )  , {\bf L} , {\boldsymbol \xi }^{\sf G}_{t} )$};
\end{scope}

\begin{scope}[xshift=0.2cm]
\draw[thick, to-to] (-1.5,0) -- (1.5,0);
\node[above] at (0,0) {replacing};
\node[below] at (0,0) {$( \Phi ,{\boldsymbol \xi _{t}} )\leftrightarrow ({\sf G}^\dagger(\Phi ),{\boldsymbol \xi }^{\sf G^\dagger}_{t})$};

\draw[thick, to-to] (-1.5,-3) -- (1.5,-3);
\node[above] at (0,-3) {replacing};
\node[below] at (0,-3) {$({\sf G}(\Phi ),{\boldsymbol \xi }^{\sf G}_{t} ) \leftrightarrow( \Phi ,{\boldsymbol \xi _{t}} )$};
\end{scope}

\begin{scope}[xshift=0.4cm]
\node[above] at (5,1.5) {Interacting theory};

\node [above]at (5,0) {$\displaystyle
\int_{0}^{1} dt \, \langle {\boldsymbol \pi } \,  {\boldsymbol \xi }^{\sf G^\dagger}_{t} \big( e^{\wedge {\sf G^\dagger}(t\Phi) } \big) , {\boldsymbol \pi } \, {\bf Q} \big( e^{\wedge {\sf G^\dagger}(t\Phi) } \big) \rangle $};
\node [below]at (5,0.1) {$\displaystyle({\sf G}^\dagger(\Phi ) , {\bf Q} , {\boldsymbol \xi }^{\sf G^\dagger}_{t} )$};

\draw[thick, -to] (4.9,-0.8) -- (4.9,-1.8);
\draw[thick, to-] (5.1,-0.8) -- (5.1,-1.8);
\node [left] at (4.9,-1.2) {$\mathbf G$};
\node [right] at (5.1,-1.2) {$\mathbf G^\dagger$};

\node [above]at (5,-3) {$\displaystyle
\int_{0}^{1} dt \, \langle {\boldsymbol \pi } \,  {\boldsymbol \xi }_{t} \big( e^{\wedge t\Phi } \big) , {\boldsymbol \pi } \, {\bf L} \big( e^{\wedge t\Phi } \big) \rangle $};
\node [below]at (5,-3) {$\displaystyle( \Phi , {\bf L} , {\boldsymbol \xi _{t}} )$};
\end{scope}

\end{tikzpicture}
\end{center}
%\caption{}
%\end{figure}

\subsection{WZW-like action}

The WZW-like action can be transformed into the same forms as the $L_\infty$-type action which we discussed in the previous subsection.
In what follows, we denote the linear maps $\mathcal E_V$ and $\int_0^1 d\tau \mathcal E_V^\dagger{\scriptstyle[\tau]}$ as $\mathcal E$ and  $({\textstyle \int} \mathcal E^\dagger)$ for the notational simplicity.
Utilizing $\Psi_\mathbb X =\mathcal E({\textstyle \int} \mathcal E^\dagger) \mathbb X V$, the WZW-like action can be transform as follows\footnote{
Although (\ref{action WZW 01}) is the same form as the second line of (\ref{action_eks_form1}),
note that we can not naively identify
$\boldsymbol\pi\mathbf G^\dagger (\boldsymbol \xi_t e^{\wedge\Phi(t)})$ with $(\int \mathcal E)\partial_t V$
and
$\boldsymbol\pi\mathbf G^\dagger (e^{\wedge\Phi(t)})$ with $(\int \mathcal E)\eta V$ as states.
As we have seen in section 4,
they should be considered as the identification of representatives of certain equivalent class.
}:
\begin{align}
S_{\scriptscriptstyle \rm WZW}
&=\int^1_0 dt \langle \Psi_{\partial_t}  , Q_{\mathcal G}\Psi_\eta \rangle\no
&=\int^1_0 dt \langle  ({\textstyle \int} \mathcal E^\dagger) \partial_t V , Q ({\textstyle \int} \mathcal E^\dagger) \eta V\rangle
\label{action WZW 01}
\end{align}
Under $\xi V=0$, the action can be transformed further:
\begin{align}
S_{\scriptscriptstyle \rm WZW}
&=\int^1_0 dt \langle  ({\textstyle \int} \mathcal E^\dagger) \partial_t \xi\eta V , Q ({\textstyle \int} \mathcal E^\dagger) \eta V\rangle\no
&=\int^1_0 dt \langle  \xi_t^{({\scriptstyle \int} \mathcal E^\dagger)}  ({\textstyle \int} \mathcal E^\dagger) \eta V , Q ({\textstyle \int} \mathcal E^\dagger) \eta V\rangle.
\end{align}
where we define the deformed $\xi_t$ by
$\xi_t^{\mathsf f} := \mathsf f \, \xi_t \,\mathsf f^{-1}=\mathsf f \,  \partial_t \xi \,\mathsf f^{-1}$.
Thus, under the condition $\xi V =0$, we can represent WZW-like action in the same form as (\ref{action_eks_form1})
and it is {\it formally} described by 
$\displaystyle\big(({\textstyle \int} \mathcal E^\dagger) \eta V, Q , \xi_t^{({\scriptstyle \int} \mathcal E^\dagger)} \big)$.
Note that $({\textstyle \int} \mathcal E^\dagger)^{-1}$ is not equal to $({\textstyle \int} \mathcal E^\dagger)^\dagger$,
which is the obstacle to the compatibility of the $L_\infty$-algebra and the cyclicity.
Therefore the theory does not have an cyclic $L_\infty$-algebra naively, and in this sense the description of the action by the triplet is {\it formal}.
The non-compatibility of the cyclicity and the $L_\infty$-algebra in the large-space formulation is discussed also in \cite{Erler:2015uba}.

Under the condition $\xi V=0$, the associated string fields can be transformed as follows:
\begin{align}
\Psi_{\mathbb X} 
&= \xi_{\mathbb X}^{\mathcal E({\scriptstyle \int} \mathcal E^\dagger)}\Psi_\eta.
\end{align}
Utilizing it, one can find the WZW-like action can be described also by   
$\displaystyle\big( \Psi_\eta , Q_{\mathcal G} , \xi _{t}^{\mathcal E({\scriptstyle \int} \mathcal E^\dagger)} \big)$:
\begin{align}
S_{\scriptscriptstyle \rm WZW}
&=\int^1_0 dt \langle  \xi_t^{\mathcal E({\scriptstyle \int} \mathcal E^\dagger)} \Psi_\eta  , Q_{\mathcal G}\Psi_\eta \rangle.
\end{align}

As in the case of the small-space formulation, 
the interacting theory in the large-space formulation 
can be obtained from the free theory by {\it replacing} the field $\eta V $ and ${\boldsymbol \xi }_{t}$.
It can be understood by the following table.

%\begin{figure}[h]
\begin{center}
\begin{tikzpicture}[xscale =1 , yscale =0.9]

\begin{scope}[xshift=0.1cm]
\node[above] at (-5,1.5) {Free theory};

\node [above]at (-5,0) {$\displaystyle
\int_{0}^{1} dt \, \langle \xi_{t} \eta V , \,   Q \eta V \rangle$};
\node [below] at (-5,0) {$\displaystyle( \eta V, Q ,  \xi _{t} )$};

\draw[thick, -to] (-5.1,-0.8) -- (-5.1,-1.8);
\draw[thick, to-] (-4.9,-0.8) -- (-4.9,-1.8);
\node [left] at (-5.1,-1.2) {$\mathcal E$};
\node [right] at (-4.9,-1.2) {$\mathcal E^\dagger$};

\node [above]at (-5,-3) {$\displaystyle
\int_{0}^{1} dt \, \langle \xi_{t}^{\mathcal E} \mathcal E( \eta V) , \,   Q_{\mathcal G} \mathcal E (\eta V) \rangle$};
\node [below] at (-5,-3) {$\displaystyle \big( \mathcal E  (\eta V) , Q_{\mathcal G}, \xi _{t}^{\mathcal E} \big)$};
\end{scope}

\draw[thick, to-to] (-2.5,0) -- (2.5,0);
\node[above] at (0,0) {replacing};
\node[below] at (0,0) {$( \eta V , \xi _{t} )\leftrightarrow \big(\textstyle \int {\cal E}^\dagger(\eta V), \xi^{\int \mathcal E^\dagger}_{t}\big)$};

\draw[thick, to-to] (-2.5,-3) -- (2.5,-3);
\node[above] at (0,-3) {replacing};
\node[below] at (0,-3) {$\big( \mathcal E  (\eta V) , \xi _{t}^{\mathcal E} \big)\leftrightarrow 
(\Psi_\eta, \xi^{\mathcal E\int \mathcal E^\dagger}_{t})
$};

\begin{scope}[xshift=0.7cm]
\node[above] at (5,1.5) {Interacting theory};

\node [above]at (5,0) {$\displaystyle
\int^1_0 dt \langle  \xi_t^{({\scriptstyle \int} \mathcal E^\dagger)}  ({\textstyle \int} \mathcal E^\dagger) \eta V , Q ({\textstyle \int} \mathcal E^\dagger) \eta V\rangle
$};
\node [below]at (5,0.15) {$\displaystyle\big(({\textstyle \int} \mathcal E^\dagger) \eta V, Q , \xi_t^{({\scriptstyle \int} \mathcal E^\dagger)} \big)$};

\draw[thick, -to] (4.9,-0.8) -- (4.9,-1.8);
\draw[thick, to-] (5.1,-0.8) -- (5.1,-1.8);
\node [left] at (4.9,-1.2) {$\mathcal E$};
\node [right] at (5.1,-1.2) {$\mathcal E^\dagger$};

\node [above]at (5,-3.1) {$\displaystyle
\int^1_0 dt \langle  \xi_t^{\mathcal E({\scriptstyle \int} \mathcal E^\dagger)} \Psi_\eta  , Q_{\mathcal G}\Psi_\eta \rangle
$};
\node [below]at (5,-3) {$\displaystyle\big( \Psi_\eta , Q_{\mathcal G} , \xi _{t}^{\mathcal E({\scriptstyle \int} \mathcal E^\dagger)} \big)$};
\end{scope}

\end{tikzpicture}
\end{center}
%\caption{}
%\end{figure}

%%%%%%%%%%%%%%%%%%%%%%%%%%%%%%%%%%%%%%%%%%%%%%%%%%%%%%%%%%%%%%%%%%%%%%%%%%%


\small

\begin{thebibliography}{99}
 
%Bosonic String field theory

%\cite{Witten:1985cc}
\bibitem{Witten:1985cc}
  E.~Witten,
  ``Noncommutative Geometry and String Field Theory,''
  Nucl.\ Phys.\  B {\bf 268}, 253 (1986).
  %%CITATION = NUPHA,B268,253;%%

%\cite{Hata:1986}
\bibitem{Hata:1986}
  H.~Hata, K.~Itoh, T.~Kugo, H.~Kunitomo and K.~Ogawa,
  ``Covariant String Field Theory,''
  Phys.\ Rev.\  D {\bf 34} (1986) 2360, %. 
%  H.~Hata, K.~Itoh, T.~Kugo, H.~Kunitomo and K.~Ogawa,
  ``LOOP AMPLITUDES IN COVARIANT STRING FIELD THEORY,''
  Phys.\ Rev.\  D {\bf 35} (1987) 1356.
  %%CITATION = PHRVA,D35,1356;%%

    %\cite{Thorn:1986qj}
\bibitem{Thorn:1986qj}
  C.~B.~Thorn,
  ``Perturbation Theory for Quantized String Fields,''
  Nucl.\ Phys.\ B {\bf 287} (1987) 61.
  %%CITATION = NUPHA,B287,61;%%
  %131 citations counted in INSPIRE as of 29 Dec 2014



%\cite{Kugo:1989}
\bibitem{Kugo:1989}
  T.~Kugo, H.~Kunitomo and K.~Suehiro,
  ``Nonpolynomial Closed String Field Theory,''
  Phys.\ Lett.\  B {\bf 226}, 48 (1989). 
  T.~Kugo and K.~Suehiro,
  ``NONPOLYNOMIAL CLOSED STRING FIELD THEORY: ACTION AND ITS GAUGE INVARIANCE,''
  Nucl.\ Phys.\  B {\bf 337}, 434 (1990).
  %%CITATION = NUPHA,B337,434;%%


%\cite{Sonoda:1989wa}
\bibitem{Sonoda:1989wa}
  H.~Sonoda and B.~Zwiebach,
  ``Closed String Field Theory Loops With Symmetric Factorizable Quadratic Differentials,''
  Nucl.\ Phys.\ B {\bf 331} (1990) 592.
  %%CITATION = NUPHA,B331,592;%%
  %46 citations counted in INSPIRE as of 29 Dec 2014

%\cite{Sen:1990ff}
\bibitem{Sen:1990ff} 
  A.~Sen,
  ``Equations Of Motion In Nonpolynomial Closed String Field Theory And Conformal Invariance Of Two-dimensional Field Theories,'' 
   Phys.\ Lett.\ B {\bf 241}, 350 (1990).
  %%CITATION = PHLTA,B241,350;%%

%\cite{Schubert:1991en}
\bibitem{Schubert:1991en} 
  C.~Schubert,
  ``The Finite gauge transformations in closed string field theory,''  Lett.\ Math.\ Phys.\  {\bf 26}, 259 (1992).
  %%CITATION = LMPHD,26,259;%%

%\cite{Zwiebach:1992ie}
\bibitem{Zwiebach:1992ie}
  B.~Zwiebach,
  ``Closed string field theory: Quantum action and the B-V master equation,''
  Nucl.\ Phys.\  B {\bf 390}, 33 (1993)
  [arXiv:hep-th/9206084].
  %%CITATION = NUPHA,B390,33;%%


%\cite{Gaberdiel:1997ia}
\bibitem{Gaberdiel:1997ia}
  M.~R.~Gaberdiel and B.~Zwiebach,
  ``Tensor constructions of open string theories. 1: Foundations,''
  Nucl.\ Phys.\  B {\bf 505}, 569 (1997)
  [arXiv:hep-th/9705038].
  %%CITATION = NUPHA,B505,569;%%

 %\cite{Nakatsu:2001da}
\bibitem{Nakatsu:2001da}
  T.~Nakatsu,
  ``Classical open string field theory: A(infinity) algebra, renormalization group and boundary states,''
  Nucl.\ Phys.\ B {\bf 642} (2002) 13
  [hep-th/0105272].
  %%CITATION = HEP-TH/0105272;%%
  %24 citations counted in INSPIRE as of 29 Dec 2014

 


%Superstring fields
%old-fashion

%\cite{Witten:1986qs}
\bibitem{Witten:1986qs}
  E.~Witten,
  ``Interacting Field Theory of Open Superstrings,''
  Nucl.\ Phys.\ B {\bf 276} (1986) 291.
  %%CITATION = NUPHA,B276,291;%%
  %449 citations counted in INSPIRE as of 29 Jul 2014

%\cite{Wendt:1987zh}
\bibitem{Wendt:1987zh}
  C.~Wendt,
  ``Scattering Amplitudes and Contact Interactions in Witten's Superstring Field Theory,''
  Nucl.\ Phys.\ B {\bf 314} (1989) 209.
  %%CITATION = NUPHA,B314,209;%%
  %83 citations counted in INSPIRE as of 29 Jul 2014

%\cite{Arefeva:1989cp}
\bibitem{Arefeva:1989cp}
  I.~Y.~Arefeva, P.~B.~Medvedev and A.~P.~Zubarev,
  ``New Representation For String Field Solves The Consistency Problem For Open Superstring Field Theory,''
  Nucl.\ Phys.\ B {\bf 341} (1990) 464.
  %%CITATION = NUPHA,B341,464;%%  %103 citations counted in INSPIRE as of 16 Mar 2013

%\cite{Preitschopf:1989fc}
\bibitem{Preitschopf:1989fc}
  C.~R.~Preitschopf, C.~B.~Thorn and S.~A.~Yost,
  ``Superstring Field Theory,''
  Nucl.\ Phys.\ B {\bf 337} (1990) 363.
  %%CITATION = NUPHA,B337,363;%%  %110 citations counted in INSPIRE as of 25 Mar 2013


%\cite{Saroja:1992vw}
\bibitem{Saroja:1992vw}
  R.~Saroja and A.~Sen,
  ``Picture changing operators in closed fermionic string field theory,''
  Phys.\ Lett.\ B {\bf 286} (1992) 256
  [hep-th/9202087].
  %%CITATION = HEP-TH/9202087;%%
  %14 citations counted in INSPIRE as of 26 Jul 2014

%\cite{Jurco:2013qra}
\bibitem{Jurco:2013qra}
  B.~Jurco and K.~Muenster,
  ``Type II Superstring Field Theory: Geometric Approach and Operadic Description,''
  JHEP {\bf 1304} (2013) 126
  [arXiv:1303.2323 [hep-th]].
  %%CITATION = ARXIV:1303.2323;%%
  %7 citations counted in INSPIRE as of 29 Jul 2014






%the large Hilbert space description 

%\cite{Berkovits:1995ab}
\bibitem{Berkovits:1995ab}
  N.~Berkovits,
  ``SuperPoincare invariant superstring field theory,''
  Nucl.\ Phys.\ B {\bf 450} (1995) 90
   [Erratum-ibid.\ B {\bf 459} (1996) 439]
  [hep-th/9503099].
  %%CITATION = HEP-TH/9503099;%%
  %184 citations counted in INSPIRE as of 26 Jul 2014

%\cite{Berkovits:1999bs}
\bibitem{Berkovits:1999bs}
  N.~Berkovits and C.~T.~Echevarria,
  ``Four point amplitude from open superstring field theory,''
  Phys.\ Lett.\ B {\bf 478} (2000) 343
  [hep-th/9912120].
  %%CITATION = HEP-TH/9912120;%%
  %36 citations counted in INSPIRE as of 29 Jul 2014


%\cite{Berkovits:1998bt}
\bibitem{Berkovits:1998bt}
  N.~Berkovits,
  ``A New approach to superstring field theory,''
  Fortsch.\ Phys.\  {\bf 48} (2000) 31
  [hep-th/9912121].
  %%CITATION = HEP-TH/9912121;%%
  %56 citations counted in INSPIRE as of 26 Jul 2014

%\cite{Berkovits:2004xh}
\bibitem{Berkovits:2004xh}
  N.~Berkovits, Y.~Okawa and B.~Zwiebach,
  ``WZW-like action for heterotic string field theory,''
  JHEP {\bf 0411} (2004) 038
  [hep-th/0409018], %.
  %%CITATION = HEP-TH/0409018;%%
  %27 citations counted in INSPIRE as of 26 Jul 2014
%
%\cite{Okawa:2004ii}
%\bibitem{Okawa:2004ii}
  Y.~Okawa and B.~Zwiebach,
  ``Heterotic string field theory,''
  JHEP {\bf 0407} (2004) 042
  [hep-th/0406212].
  %%CITATION = HEP-TH/0406212;%%
  %21 citations counted in INSPIRE as of 26 Jul 2014

%\cite{Iimori:2013kha}
\bibitem{Iimori:2013kha}
  Y.~Iimori, T.~Noumi, Y.~Okawa and S.~Torii,
  ``From the Berkovits formulation to the Witten formulation in open superstring field theory,''
  JHEP {\bf 1403} (2014) 044
  [arXiv:1312.1677 [hep-th]].
  %%CITATION = ARXIV:1312.1677;%%
  %2 citations counted in INSPIRE as of 26 Jul 201


%\cite{Matsunaga:2013mba}
\bibitem{Matsunaga:2013mba}
  H.~Matsunaga,
  ``Construction of a Gauge-Invariant Action for Type II Superstring Field Theory,''
  arXiv:1305.3893 [hep-th]. 
  %%CITATION = ARXIV:1305.3893;%%
  %6 citations counted in INSPIRE as of 26 Jul 2014

%\cite{Matsunaga:2014wpa}
\bibitem{Matsunaga:2014wpa}
  H.~Matsunaga,
  ``Nonlinear gauge invariance and WZW-like action for NS-NS superstring field theory,''
  arXiv:1407.8485 [hep-th].
  %%CITATION = ARXIV:1407.8485;%%

%EKS
%\cite{Erler:2013xta}
\bibitem{Erler:2013xta}
  T.~Erler, S.~Konopka and I.~Sachs,
  ``Resolving Witten`s superstring field theory,''
  JHEP {\bf 1404} (2014) 150
  [arXiv:1312.2948 [hep-th]].
  %%CITATION = ARXIV:1312.2948;%%
  %3 citations counted in INSPIRE as of 26 Jul 2014
   
%\cite{Erler:2014eba}
\bibitem{Erler:2014eba}
  T.~Erler, S.~Konopka and I.~Sachs,
  ``NS-NS Sector of Closed Superstring Field Theory,''
  arXiv:1403.0940 [hep-th].
  %%CITATION = ARXIV:1403.0940;%%
  %2 citations counted in INSPIRE as of 26 Jul 2014

%Homotopy Algebras

%\cite{Lada:1992wc}
\bibitem{Lada:1992wc} 
  T.~Lada and J.~Stasheff,
  ``Introduction to SH Lie algebras for physicists,''
  Int.\ J.\ Theor.\ Phys.\  {\bf 32}, 1087 (1993)
  [hep-th/9209099].
  %%CITATION = HEP-TH/9209099;%%
  %132 citations counted in INSPIRE as of 18 juin 2015
  
%\cite{Kajiura:2001ng}
\bibitem{Kajiura:2001ng}
  H.~Kajiura,
  ``Homotopy algebra morphism and geometry of classical string field theory,''
  Nucl.\ Phys.\ B {\bf 630} (2002) 361
  [hep-th/0112228].
  %%CITATION = HEP-TH/0112228;%%
  %23 citations counted in INSPIRE as of 29 Dec 2014
  %\cite{Kajiura:2003ax}
%\bibitem{Kajiura:2003ax}
%  H.~Kajiura,
  ``Noncommutative homotopy algebras associated with open strings,''
  Rev.\ Math.\ Phys.\  {\bf 19} (2007) 1
  [math/0306332 [math-qa]].
  %%CITATION = MATH/0306332;%%
  %20 citations counted in INSPIRE as of 29 Dec 2014

%%\cite{Getzler:2007}
\bibitem{Getzler:2007}
  E.~Getzler,
  ``LIE THEORY FOR NILPOTENT $L_{\infty }$-ALGEBRAS''
  [arXiv:math/0404003].
 %%

\bibitem{Ph.Dhm}
  H.~Matsunaga, ``Wess-Zumino-Witten-type formulation for NS-NS superstring field theory,'' Ph.D thesis, University of Tokyo, Japan. 

%\cite{Erler:2015rra}
\bibitem{Erler:2015rra} 
  Y.~Okawa, ``The $A_\infty$ structure from the Berkovits formulation of open superstring field theory,'' 
  Talk presented at the conference String Field Theory 2015 nara, Nara woman University, Nara, Japan. 
  T.~Erler, Y.~Okawa and T.~Takezaki,
  ``$A_\infty$ structure from the Berkovits formulation of open superstring field theory,''
  arXiv:1505.01659 [hep-th].
  %%CITATION = ARXIV:1505.01659;%%

%Erler

%\cite{Erler:2015uba}
\bibitem{Erler:2015uba} 
  T.~Erler,
  ``Relating Berkovits and $A_\infty$ Superstring Field Theories; Small Hilbert Space Perspective,''
  arXiv:1505.02069 [hep-th].
  %%CITATION = ARXIV:1505.02069;%%

%BV quantization

%\cite{Batalin:1981jr}
\bibitem{Batalin:1981jr}
  I.~A.~Batalin and G.~A.~Vilkovisky,
  ``Gauge Algebra and Quantization,''
  Phys.\ Lett.\ B {\bf 102} (1981) 27.
  %%CITATION = PHLTA,B102,27;%%
  %854 citations counted in INSPIRE as of 29 Dec 2014
%\cite{Batalin:1984jr}
%\bibitem{Batalin:1984jr}
%  I.~A.~Batalin and G.~A.~Vilkovisky,
  ``Quantization of Gauge Theories with Linearly Dependent Generators,''
  Phys.\ Rev.\ D {\bf 28} (1983) 2567
   [Erratum-ibid.\ D {\bf 30} (1984) 508].
  %%CITATION = PHRVA,D28,2567;%%
  %930 citations counted in INSPIRE as of 29 Dec 2014

%seeking for gauge fixing
  
  %\cite{Torii:2011zz}
\bibitem{Torii:2011}
  S.~Torii,
  ``Gauge fixing of open superstring field theory in the Berkovits non-polynomial formulation,''
  Prog.\ Theor.\ Phys.\ Suppl.\  {\bf 188} (2011) 272
  [arXiv:1201.1763 [hep-th]], 
  %%CITATION = ARXIV:1201.1763;%%
  %7 citations counted in INSPIRE as of 26 Jul 2014
%\cite{Torii:2012nj}
%\bibitem{Torii:2012nj}
  %S.~Torii,
  ``Validity of Gauge-Fixing Conditions and the Structure of Propagators in Open Superstring Field Theory,''
  JHEP {\bf 1204} (2012) 050
  [arXiv:1201.1762 [hep-th]]. %.
  %%CITATION = ARXIV:1201.1762;%%
  %8 citations counted in INSPIRE as of 26 Jul 2014

%\cite{Kroyter:2012ni}
\bibitem{Kroyter:2012ni}
  M.~Kroyter, Y.~Okawa, M.~Schnabl, S.~Torii and B.~Zwiebach,
  ``Open superstring field theory I: gauge fixing, ghost structure, and propagator,''
  JHEP {\bf 1203} (2012) 030
  [arXiv:1201.1761 [hep-th]].
  %%CITATION = ARXIV:1201.1761;%%
  %14 citations counted in INSPIRE as of 26 Jul 2014

%\cite{Berkovits:2012np}
\bibitem{Berkovits:2012np} 
  N.~Berkovits,
  ``Constrained BV Description of String Field Theory,''
  JHEP {\bf 1203}, 012 (2012)
  [arXiv:1201.1769 [hep-th]].
  %%CITATION = ARXIV:1201.1769;%%
  %9 citations counted in INSPIRE as of 18 Jun 2015

%R sector 

%\cite{Berkovits:2001im}
\bibitem{Berkovits:2001im}
  N.~Berkovits,
  ``The Ramond sector of open superstring field theory,''
  JHEP {\bf 0111} (2001) 047
  [hep-th/0109100].
  %%CITATION = HEP-TH/0109100;%%
  %45 citations counted in INSPIRE as of 26 Jul 2014

%\cite{Michishita:2004by}
\bibitem{Michishita:2004by}
  Y.~Michishita,
  ``A Covariant action with a constraint and Feynman rules for fermions in open superstring field theory,''
  JHEP {\bf 0501} (2005) 012
  [hep-th/0412215].
  %%CITATION = HEP-TH/0412215;%%
  %20 citations counted in INSPIRE as of 29 Jul 2014

%\cite{Kunitomo:2013mqa}
\bibitem{Kunitomo:2013mqa}
  H.~Kunitomo,
  ``The Ramond Sector of Heterotic String Field Theory,''
  PTEP {\bf 2014} (2014) 4,  043B01
  [arXiv:1312.7197 [hep-th]], 
  %%CITATION = ARXIV:1312.7197;%%
  %3 citations counted in INSPIRE as of 26 Jul 2014
%
%\cite{Kunitomo:2014hba}
%\bibitem{Kunitomo:2014hba}
%  H.~Kunitomo,
  ``First-Order Equations of Motion for Heterotic String Field Theory,''
  arXiv:1407.0801 [hep-th].
  %%CITATION = ARXIV:1407.0801;%

%\cite{Kunitomo:2014qla}
\bibitem{Kunitomo:2014qla} 
  H.~Kunitomo,
  ``Symmetries and Feynman rules for the Ramond sector in open superstring field theory,''
  PTEP {\bf 2015}, no. 3, 033B11 (2015)
  [arXiv:1412.5281 [hep-th]].
  %%CITATION = ARXIV:1412.5281;%%
  %1 citations counted in INSPIRE as of 11 juin 2015

\bibitem{Erler:talk}
  T.~Erler, 
  ``Ramond Equations of Motion in Superstring Field Theory,''
  Talk presented at the conference String Field Theory and Related Aspects VII, Sichuan University, Chengdu, China. 
  
\end{thebibliography}
\end{document}